\documentclass[12pt]{article}
\usepackage{amsmath}
\usepackage{amsfonts}
\usepackage{amssymb}
\usepackage{graphicx}
\graphicspath{{.//}}
\usepackage{geometry}
\usepackage{setspace}
\usepackage{comment}
\usepackage{mathrsfs}
\usepackage{xcolor}
\usepackage{bm}
\usepackage{natbib}
\usepackage{placeins}
\usepackage{booktabs}
\usepackage{pdflscape}
\usepackage{authblk}
\usepackage{rotating}

\title{Inference for Error-Prone Count Data: Estimation under a Binomial Convolution Framework}
\author[1]{Yuqiu Yang}
\author[2]{Christina Vu}
\author[2,3]{Cornelis J. Potgieter}
\author[4]{Xinlei Wang}
\author[5]{Akihito Kamata}

\affil[1]{UT Southwestern Medical Center}
\affil[2]{Texas Christian University}
\affil[3]{University of Johannesburg}
\affil[4]{University of Texas at Arlington}
\affil[5]{Southern Methodist University}

\date{}

\begin{document}

\maketitle

\begin{abstract}
    Measurement error in count data is common but underexplored in the literature, particularly in contexts where observed scores are bounded and arise from discrete scoring processes. Motivated by applications in oral reading fluency assessment, we propose a binomial convolution framework that extends binary misclassification models to settings where only the aggregate number of correct responses is observed, and errors may involve both overcounting and undercounting the number of events. The model accommodates distinct true positive and true negative accuracy rates and preserves the bounded nature of the data.

    Assuming the availability of both contaminated and error-free scores on a subset of items, we develop and compare three estimation strategies: maximum likelihood estimation (MLE), linear regression, and generalized method of moments (GMM). Extensive simulations show that MLE is most accurate when the model is correctly specified but is computationally intensive and less robust to misspecification. Regression is simple and stable but less precise, while GMM offers a compromise in model dependence, though it is sensitive to outliers.

    In practice, this framework supports improved inference in unsupervised settings where contaminated scores serve as inputs to downstream analyses. By quantifying accuracy rates, the model enables score corrections even when no specific outcome is yet defined. We demonstrate its utility using real oral reading fluency data, comparing human and AI-generated scores. Findings highlight the practical implications of estimator choice and underscore the importance of explicitly modeling asymmetric measurement error in count data.
\end{abstract}

\doublespacing

\section{Introduction}

Measurement error in statistical modeling has been extensively studied for continuous variables, with foundational work documenting its effects on bias and efficiency in parameter estimation, as well as power loss in statistical inference \citep{carroll1995measurement, fuller2009measurement}. In contrast, the consequences of measurement error in count data remain comparatively underexplored, despite the widespread use of discrete measurements in fields such as education, public health, and epidemiology. From a statistical modeling perspective, there is no widely accepted framework for how measurement error should be formulated in count data, in part because the additive noise assumption that underpin classical measurement error models are incompatible with the discrete and often bounded nature of count outcomes.

This paper is motivated by oral reading fluency (ORF) assessments in elementary education, where the number of words correctly read aloud by a student serves as a key measure of early literacy \citep{fuchs2001oral}. These counts are typically generated by either a human rater or an automated scoring system, both of which are susceptible to error. Human raters may miscount due to distractions, inconsistent scoring practices, or real-time processing demands. Automated systems, while offering increased efficiency, can introduce misclassification due to poor audio quality, background noise, or limitations in speech recognition algorithms. Empirical comparisons suggest that automatic speech recognition (ASR) scoring methods yield scores broadly comparable to human scoring, though discrepancies in word-level agreement rates remain \citep{nese2021evidence}. To evaluate the extent of measurement error, we consider a subset of ORF data with triplets $(X, Y_1, Y_2)$, where $X$ denotes the true word count established by multi-rater consensus, and $Y_1$ and $Y_2$ are scores obtained via human and automated scoring, respectively. This curated subset allows for direct comparison of observed scores to a reliable benchmark. However, in practical applications, such true scores are rarely available; for most students in the larger motivating dataset, only a single observed score, typically $Y_2$, is recorded. This limitation underscores the need for statistical models that can estimate and adjust for measurement error in the absence of verified ground truth.

Misclassification, a specific form of measurement error, has been widely studied in binary and categorical settings. In these contexts, misclassification probabilities are often modeled via matrices $\pi_{xy} = P(Y = y \, | \, X = x)$ for discrete support sets $\mathcal{X}, \mathcal{Y}$ \citep[Section 3.7]{grace2016statistical}. While such frameworks can, in principle, be extended to count data, they become impractical when the outcome space is large and observed $(x, y)$ combinations are sparse. As a result, measurement error in count data is often addressed using methods developed for continuous outcomes, assuming approximately unbounded or normally distributed error \citep[e.g.,][]{yan2016class}. Although such approximations are sometimes defensible, they fail to respect the bounded and integer-valued structure of many count-based outcomes.

Several methods have been proposed to address measurement error in discrete settings. \citet{wu2015bayesian} introduced a Bayesian model for misclassification in Poisson outcomes, though their approach assumes unbounded support. The SIMEX (simulation-extrapolation) algorithm \citep{cook1994simulation}, extended by \citet{kuchenhoff2006general} to misclassified categorical predictors, is primarily designed for correcting predictor error in generalized linear models, rather than modeling the measurement error mechanism itself. A limited number of studies have tackled misclassification in count outcomes directly. \citet{whittemore1991poisson} incorporated supplemental data to adjust for misclassified cancer mortality counts, while \citet{mwalili2008zero} extended these ideas to overdispersed and zero-inflated settings. \citet{muff2018bias} demonstrated the substantive impact of miscounted outcomes in a clinical trial reanalysis, and more recently, \citet{zhang2023zero} proposed a Bayesian framework for zero-inflated Poisson models with response error. \citet{hamada2022analyzing} offered a latent variable model that indirectly adjusts for systematic under- or over-counting, though their approach does not explicitly model the contamination mechanism.

While valuable, these existing methods generally treat measurement error as a nuisance in regression modeling, where the mismeasured variable appears as a covariate or response. These approaches are typically designed for supervised settings with a clearly defined outcome variable. In contrast, our focus lies in modeling the measurement error process directly, with no observed outcome of interest. The count variable under consideration is either a predictor or a variable of substantive importance in downstream analyses, but not itself the response. Consequently, approaches like SIMEX are inapplicable here, as they rely on outcome-driven correction mechanisms.

To address this gap, we develop a binomial convolution model that explicitly captures structured misclassification in bounded discrete data. Our framework respects the count variable’s finite support and enables valid inference on the measurement process itself. In the context of ORF assessments, this allows us to quantify discrepancies between human and machine-generated scores without requiring a downstream outcome model. More broadly, the proposed methodology contributes to the measurement error literature by providing a flexible and interpretable approach for count variables in settings akin to unsupervised learning, where the variable of interest is observed with error but not conditioned on an external response.

The remainder of the paper is organized as follows. Section 2 introduces the binomial convolution framework and the three proposed estimators. Section 3 addresses standard error estimation. Section 4 presents simulation results. Section 5 applies the methods to oral reading fluency data. Section 6 concludes with a discussion.

\section{Methodology}

\subsection{Model Formulation} \label{sec:Model_form}

Let $X\in \{0,1,\ldots,N\}$ denote the true number of successes in a finite collection of $N$ trials. We remain agnostic regarding the true distribution of $X$, acknowledging that the composite trials constituting the count $X$ need not be independent and may have varying success probabilities. Furthermore, we let $\mu_x=\mathrm{E}[X]$ and $\sigma_x^2=\mathrm{Var}[X]$ denote its mean and variance, respectively, and define the expected success proportion as $p_x = \mu_x / N$.  

Similarly, let $Y\in \{0,1,\ldots,N\}$ denote a misclassified or measurement-contaminated count. Define its mean, variance, and expected success proportion analogously as $\mu_y$, $\sigma_y^2$, and $p_y$, respectively. Let $\sigma_{xy}=\mathrm{Cov}[X,Y]$ denote the covariance between the true and observed success counts. We assume the availability of paired observations $(X, Y)$, meaning the true and error-prone counts are jointly recorded.

To model the measurement error present in $Y$, we decompose it into two components: the true positive count, $\mathrm{TP}$, representing correctly classified successes, and the false positive count, $\mathrm{FP}$, representing failures misclassified as successes. Conditional on $X$, we assume
\begin{equation}
    \mathrm{TP} \, | \, X \sim \text{Bin}(X, \pi_{\text{tp}}) \quad\text{and}\quad \mathrm{FP} \, | \, X \sim \text{Bin}(N - X, 1 - \pi_{\text{tn}}),
\end{equation}
where $\pi_{\text{tp}}$ denotes the true positive rate (sensitivity), and $\pi_{\text{tn}}$ the true negative rate (specificity). Since $\mathrm{TP}$ and $\mathrm{FP}$ originate from disjoint subsets of the $N$ trials, namely the $X$ successes and the $N - X$ failures, we assume they are conditionally independent given $X$; that is, $\mathrm{TP} \perp \mathrm{FP} \, | \, X$. The error-prone count is then expressed as $Y = \mathrm{TP} + \mathrm{FP}$.

The conditional distribution $Y \, | \, X$ is a \textit{binomial convolution}, representing the sum of two independent binomial variables with distinct success probabilities. Its conditional moments are
\begin{equation}
    \text{E}[Y | X] = X \pi_{\text{tp}} + (N - X)(1 - \pi_{\text{tn}}) \label{eq:cond_mean}
\end{equation}
and
\begin{equation}
    \text{Var}[Y|X] = X \pi_{\text{tp}} (1 - \pi_{\text{tp}}) + (N - X)(1 - \pi_{\text{tn}}) \pi_{\text{tn}}. \label{eq:cond_var}
\end{equation}
Applying the law of iterated expectations yields the marginal moments of $Y$,
\begin{equation}
    \mu_y = \mu_x \pi_{\text{tp}} + (N - \mu_x)(1 - \pi_{\text{tn}}), \label{eq:mu_y}
\end{equation}
and
\begin{eqnarray}
    \sigma_y^2 &=& \mu_x [\pi_{\text{tp}} (1- \pi_{\text{tp}}) - \pi_{\text{tn}} (1- \pi_{\text{tn}})] \notag \\
    && +\, N\pi_{\text{tn}}(1-\pi_{\text{tn}}) + \sigma_x^2 (\pi_{\text{tp}}+\pi_{\text{tn}}-1)^2. \label{eq:var_y}
\end{eqnarray}
Similarly, the covariance between the true and error-prone counts is
\begin{equation}
    \sigma_{xy} = \sigma_x^2 (\pi_{\mathrm{tp}}+\pi_{\mathrm{tn}}-1). \label{eq:cov_xy}
\end{equation}

These expressions illustrate the complex impact of measurement error in count data. In contrast to the classic additive measurement error model, which is conditionally unbiased, the expected value  $\mu_y$ may be shifted in either direction relative to the true mean $\mu_x$. Moreover, unlike the traditional model where variance is always inflated, here the variance $\sigma_y^2$ may be either inflated or attenuated depending on the misclassification parameters.

Overall, the binomial convolution model provides a flexible and realistic framework for analyzing count data subject to misclassification. It respects the bounded, discrete nature of counts and accommodates structured dependence between true and contaminated values. Estimation of the misclassification rates $(\pi_{\mathrm{tp}}, \pi_{\mathrm{tn}})$ proceeds via three approaches developed in the following subsections: a fully parametric estimator based on the full likelihood for the conditional distribution of $Y \,|\, X$, leaving the marginal distribution of $X$ unspecified; a nonparametric regression-style estimator based solely on the conditional mean of $Y \,|\, X$; and a generalized method of moments (GMM) estimator that uses both first and second unconditional moments implied by the binomial convolution model.

\subsection{Maximum Likelihood Estimation}

This section outlines the estimation of the misclassification parameters $(\pi_{\text{tp}},\pi_{\text{tn}})$ using maximum likelihood. As we do not impose a distributional model on the true count, estimation is carried out conditionally on $X$, analogous to the way regression models treat predictor variables as fixed.

Assume independent pairs $(X_1, Y_1), \ldots, (X_n, Y_n)$ are observed, where each pair $(X_j,Y_j)$ corresponds to a known number of trials $N_j$. Under the binomial convolution error model, the probability mass function (pmf) of $ Y_j $ given $ X_j $, denoted $ f_j(y | x) $, depends on both $N_j$ and the parameters $(\pi_{\text{tp}},\pi_{\text{tn}})$. For brevity, the notation suppresses parameter dependence. The conditional pmf is given by
\begin{equation}
f_j(y|x) = \sum_{k=\max(0, x + y - N_j)}^{\min(x, y)} \binom{x}{k} \binom{N_j - x}{y - k} \pi_{\text{tp}}^k (1 - \pi_{\text{tp}})^{x - k} (1 - \pi_{\text{tn}})^{y - k} \pi_{\text{tn}}^{(N_j - x) - (y - k)}
\end{equation}
for $x,y\in\{0,1,\ldots,N_j\}$. This expression represents the convolution of two independent binomial distributions. While direct evaluation is possible, it can be computationally intensive. A more efficient alternative leverages the discrete Fourier transform (DFT), yielding
\begin{equation}
f_j(y|x) = \frac{1}{N_j+1} \sum_{l=0}^{N_j} \bigl[ (1-\pi_{\text{tp}} + \pi_{\text{tp}} e^{i s})^x \, \{\pi_{\text{tn}} + (1-\pi_{\text{tn}}) e^{i s}\}^{N_j - x} \bigr] e^{-i y s},
\end{equation}
where $i=\sqrt{-1}$ is the imaginary unit, $ s = \omega l $, and $ \omega = 2\pi/(N_j+1)$. 

The full log-likelihood function for $n$ independent observations under the binomial convolution contamination model is
$\ell(\pi_{\text{tp}}, \pi_{\text{tn}}) = \sum_{j=1}^{n} \log f_j(Y_j|X_j)$. Maximization of $\ell(\pi_{\text{tp}}, \pi_{\text{tn}})$ requires numerical optimization. Since only two parameters are involved, standard routines such as quasi-Newton methods (e.g., BFGS) are effective. In our empirical applications, where the largest value of $N_j$ is approximately $70$, direct likelihood evaluation is computationally feasible. For larger $N_j$, the DFT-based representation offers a more scalable alternative.

Closed-form expressions for the first and second derivatives of the log-likelihood are available but complex, particularly under the DFT formulation. These expressions, though omitted here for clarity, are available from the authors upon request. They may be used to implement gradient-based optimization or to compute asymptotic standard errors.

Approximate standard errors for the maximum likelihood estimators (MLEs) are obtained from the observed information matrix
\begin{equation} (\text{SE} (\hat{\pi}_{\text{tp}}), \text{SE} (\hat{\pi}_{\text{tn}})) = \sqrt{\text{diag} \big( \mathcal{I}^{-1} \big)}, \end{equation}
where $\mathcal{I}$ denotes the observed information matrix, defined by the negative Hessian of the log-likelihood evaluated at the MLE. The square-root is applied element-wise. 

Joint $100(1-\alpha)\%$ confidence regions can be constructed using the likelihood ratio test, $
\mathrm{CR} = \{ (\pi_{\text{tp}}, \pi_{\text{tn}}) : 2 [ \ell(\hat{\pi}_{\text{tp}}, \hat{\pi}_{tn}) - \ell(\pi_{\text{tp}}, \pi_{\text{tn}}) ] \leq \chi^2_{2, \alpha} \}$, where $\chi^2_{2, \alpha}$ denotes the upper $\alpha$ quantile of the chi-squared distribution with $2$ degrees of freedom. Marginal confidence intervals can be constructed using profile likelihoods by fixing one parameter and maximizing over the other, with degrees of freedom adjusted accordingly.

\subsection{Linear Regression}

Recall the conditional expectation of $Y$ given $X$ in \eqref{eq:cond_mean}. To make the dependence on the number of trials explicit, we write $\text{E}[Y_j | X_j] = \pi_{\text{tp}} X_j + (1-\pi_{\text{tn}})(N_j - X_j)$. This conditional mean is linear in both $X_j$ and $N_j-X_j$, motivating a regression-based estimator. Recall the sample is $(X_j,Y_j)$, $j=1,\ldots,n$, with each unit having a known total count $N_j$. Define the regression parameter vector $\bm{\beta}^\top = [\pi_{\text{tp}},1-\pi_{\text{tn}}]$, and construct the following design and response matrices,
\begin{equation}
    \mathbf{X} = \begin{bmatrix} X_1 \\ X_2 \\ \vdots \\ X_n \end{bmatrix},\quad 
    \mathbf{D} = \begin{bmatrix}
        X_1 & N_1 - X_1 \\
        X_2 & N_2 - X_2 \\
        \vdots & \vdots \\
        X_n & N_n - X_n
    \end{bmatrix}, \quad\text{and}\quad \mathbf{Y} = \begin{bmatrix} Y_1 \\ Y_2 \\ \vdots \\ Y_n \end{bmatrix}.
\end{equation}
Since $\text{E}[\mathbf{Y}|\mathbf{X}] = \mathbf{D}\bm{\beta}$, the misclassification parameters can be estimated by ordinary least squares (OLS), yielding
\begin{equation}
    \hat{\bm{\beta}} = (\mathbf{D}^\top \mathbf{D})^{-1} \mathbf{D}^\top \mathbf{Y}. \label{eq:OLS_est}
\end{equation}
Notably, this estimator is unbiased under the assumption that the conditional mean follows the structure implied by the binomial convolution model, even if higher-order moments or distributional assumptions are misspecified.

Estimated misclassification rates are then given by
\begin{equation}
    \hat{\pi}_{\text{tp}} = \max\{1,\min\{0,\hat{\beta}_{\text{tp}}\}\} \quad\text{and}\quad \hat{\pi}_{\text{tn}} = \max\{1,\min\{0,1 - \hat{\beta}_{\text{tn}}\}\}.
\end{equation}
where the max-min operation ensures the estimates remain within the admissible interval $[0,1]$, as the unconstrained OLS estimates may fall outside.

A special case of particular interest arises when $N_j = N$ for all $j=1,\ldots,n$. In this setting, the OLS estimator in \eqref{eq:OLS_est} simplifies to
\begin{equation}
    \hat{\beta}_{\text{tp}} = \frac{\bar{Y}}{N} + \frac{S_{xy}}{S_x^2}\left(1 - \frac{\bar{X}}{N} \right) \quad\text{and}\quad
    \hat{\beta}_{\text{tn}} = \frac{\bar{Y}}{N} -  \frac{\bar{X}}{N} \frac{S_{xy}}{S_x^2}.
\end{equation}
where $\bar{X}$ and $\bar{Y}$ denote the sample means of $X$ and $Y$, $S_x^2$ is the sample variance of $X$, and $S_{xy}$ is the sample covariance between $X$ and $Y$. These expressions shed additional light on the method, revealing a direct connection between the OLS approach and moment-based estimators constructed from marginal means and second-order moments.

To assess variability, we adapt \eqref{eq:cond_var} to allow for case-specific $N_j$. The conditional variance is then $\mathrm{Var}[Y_j\,|\,X_j] = X_j\, \pi_{\mathrm{tp}}(1-\pi_{\mathrm{tp}}) + (N_j - X_j)\, \pi_{\mathrm{tn}}(1-\pi_{\mathrm{tn}})$. In matrix form, the conditional variance of the response vector can be written as $$
\mathrm{Var}[\mathbf{Y} | \mathbf{X}] = \mathrm{diag}(\mathbf{D} \bm{\alpha})\quad \text{with}\quad \bm{\alpha}^\top = [\pi_{\text{tp}}(1 - \pi_{\text{tp}}), \, \pi_{\text{tn}}(1 - \pi_{\text{tn}})].$$ From this, the conditional variance of the OLS estimator $\hat{\bm{\beta}}$ follows as  
\begin{equation}
\mathrm{Var}[\hat{\bm{\beta}} | \mathbf{X}] = (\mathbf{D}^\top \mathbf{D})^{-1} \mathbf{D}^\top \mathrm{diag}(\mathbf{D} \bm{\alpha}) \mathbf{D} (\mathbf{D}^\top \mathbf{D})^{-1}.
\end{equation}
Estimated standard errors may be obtained by substituting a consistent estimator $\hat{\bm{\alpha}}$, for example, using plug-in values based $(\hat{\pi}_{\text{tp}},\hat{\pi}_{\text{tn}})$. In the special case with $N_j = N$, an alternative approach based on the delta method and using the method-of-moments formulation can also be applied. Details are omitted.

\subsection{Generalized Method of Moments Estimation}

Building on the previous section, which introduced the regression estimator and highlighted its interpretation as a method-of-moments procedure in a special case, we now extend the framework using the Generalized Method of Moments (GMM). The moment identities derived in Section~\ref{sec:Model_form} naturally lend themselves to estimation via GMM. Unlike maximum likelihood estimation (MLE), which requires specification of the full likelihood, GMM relies only on first- and second-moment conditions. This makes it particularly attractive when the full binomial convolution model is uncertain or only partially valid. Conceptually, GMM occupies a middle ground between MLE and OLS: OLS uses only the conditional mean, MLE leverages the complete likelihood, while GMM exploits all available first- and second-order moment structure without requiring full distributional assumptions.

To introduce the GMM approach, we adopt the simplifying assumption that the number of trials is constant across observations, i.e., $N_j = N$ for all $j$. The parameters of primary interest are $\bm{\pi} = (\pi_{\text{tp}}, \pi_{\text{tn}})$, while the GMM framework additionally requires inclusion of nuisance parameters $\bm{\theta}_x = (\mu_x, \sigma_x^2)$, representing the mean and variance of the true count variable $X$. 

Based on equations \eqref{eq:mu_y}, \eqref{eq:var_y}, and \eqref{eq:cov_xy}, we define the following system of moment conditions,
\begin{align}
    g_{1j}(\bm{\pi}, \bm{\theta}_x) &= X_j - \mu_x, \notag 
 \\
    g_{2j}(\bm{\pi}, \bm{\theta}_x) &= (X_j - \mu_x)^2 - \sigma_x^2, \notag \\
    g_{3j}(\bm{\pi}, \bm{\theta}_x) &= Y_j - \mu_x \pi_{\text{tp}} - (N - \mu_x)(1 - \pi_{\text{tn}}), \notag  \\
    g_{4j}(\bm{\pi}, \bm{\theta}_x) &= (Y_j - \mu_y(\bm{\pi}, \bm{\theta}_x))^2 - \mu_x [\pi_{\text{tp}} (1 - \pi_{\text{tp}}) - \pi_{\text{tn}} (1 - \pi_{\text{tn}})] \notag \\
    &\quad - N \pi_{\text{tn}} (1 - \pi_{\text{tn}}) - \sigma_x^2 (\pi_{\text{tp}} + \pi_{\text{tn}} - 1)^2, \notag  \\
    g_{5j}(\bm{\pi}, \bm{\theta}_x) &= (X_j - \mu_x) (Y_j - \mu_y(\bm{\pi}, \bm{\theta}_x)) - \sigma_x^2 (\pi_{\text{tp}} + \pi_{\text{tn}} - 1), \notag 
\end{align}
where $\mu_y(\bm{\pi}, \bm{\theta}_x) = \mu_x \pi_{\text{tp}} + (N - \mu_x)(1 - \pi_{\text{tn}})$. Let $\bm{\theta} = (\bm{\theta}_x, \bm{\pi})$, and define the vector-valued moment function $\mathbf{g}_j(\bm{\theta}) = \big[
g_{1j}(\bm{\theta}),\ g_{2j}(\bm{\theta}),\ \cdots,\ g_{5j}(\bm{\theta})
\big]^\top.$

By construction, these moment conditions satisfy $\text{E}[\mathbf{g}_j(\bm{\theta}_0)] = \mathbf{0}$ at the true parameter vector $\bm{\theta}_0$. Estimation proceeds by equating the sample counterparts of these moments to zero and minimizing the resulting discrepancy. To this end, define the empirical moment vector
\[
\bar{\mathbf{g}}(\bm{\theta}) = \frac{1}{n} \sum_{j=1}^{n} \mathbf{g}_{j}(\bm{\theta})
\]
and the GMM objective function
\[
Q_n(\bm{\theta}) = \bar{\mathbf{g}}(\bm{\theta})^\top\, \mathbf{W}_n\, \bar{\mathbf{g}}(\bm{\theta}),
\]
where $ \mathbf{W}_n $ is a symmetric, positive definite weighting matrix. The optimal choice is $\mathbf{W}_n = n \cdot \bm{\Sigma}^{-1}$, where $\bm{\Sigma} \in \mathbb{R}^{5 \times 5}$ denotes the asymptotic covariance matrix of the scaled moment vector,
$$\bm{\Sigma} = \lim_{n \to \infty} \operatorname{Var} \left[ n^{1/2} \, \bar{\mathbf{g}}(\bm{\theta}_0) \right] = \lim_{n \to \infty} \frac{1}{n} \sum_{j=1}^{n} \operatorname{Var} \left[\, \mathbf{g}_j(\bm{\theta}_0)\, \right] = \operatorname{Var} \left[\, \mathbf{g}_1(\bm{\theta}_0)\, \right]$$
assuming independent and identically distributed observations. Any consistent estimator of $\bm{\Sigma}$ substituted in $\mathbf{W}_n$ yields a GMM estimator that is consistent and asymptotically normal. While GMM is often implemented as a two-step procedure (first obtaining preliminary parameter estimates using a simple weighting matrix, then updating the weighting matrix based on those estimates), for our application, this can be sidestepped by directly using the empirical covariance matrix $\hat{\bm{\Sigma}}$ of the moment functions. This approach provides a fully data-driven and consistent estimator of the optimal weighting matrix. The resulting GMM estimator is defined as the solution to the minimization problem
\[
\hat{\bm{\theta}} = \arg\min_{\bm{\theta}} Q_n(\bm{\theta}).
\]

Under standard conditions, the GMM estimator is consistent and asymptotically normal. Specifically,
$$n^{1/2} (\hat{\bm{\theta}} - \bm{\theta}_0) \xrightarrow{d} \text{N} \left( \mathbf{0},\ (\mathbf{G}^\top \bm{\Sigma}^{-1} \mathbf{G})^{-1} \right),$$
where $$\mathbf{G} = \left. \frac{\partial \bar{\mathbf{g}}(\bm{\theta})}{\partial \bm{\theta}} \right|_{\bm{\theta}_0}$$ is the Jacobian of the moment functions. The asymptotic variance of $\hat{\bm{\theta}}$ is consistently estimated by the sandwich formula
$$\widehat{\operatorname{Var}}(\hat{\bm{\theta}}) = (\hat{\mathbf{G}}^\top \hat{\bm{\Sigma}}^{-1} \hat{\mathbf{G}})^{-1},$$
with standard errors obtained from the square roots of the diagonal entries.

The GMM approach described above assumes a constant number of trials $N$ across all observations. In practice, however, the data may involve multiple distinct trial sizes. Let
 $\mathcal{M} = \{M_1, M_2, \ldots, M_K\}$ denote the set of unique trial sizes observed in the sample. For each $M_k \in \mathcal{M}$, define the subset of indices
$\mathcal{I}_k = \{j : N_j = M_k\}$, where $N_j$ denotes the number of trials for observation $j$. Provided that each subset $\mathcal{I}_k$ contains at least two observations, a separate GMM objective function $Q_{k}(\bm{\theta})$ can be computed for each $M_k$ using only the observations in $\mathcal{I}_k$. The overall estimator is then defined as the minimizer of the sum of these group-specific objective functions,
\[
\hat{\bm{\theta}} = \arg\min_{\bm{\theta}} \sum_{k=1}^{K} Q_{k}(\bm{\theta}),
\]
where each $Q_{k}(\bm{\theta})$ is defined analogously to the single-$N$ case but restricted to the corresponding data subset $\mathcal{I}_k$. This composite GMM formulation accommodates heterogeneity in trial sizes while preserving the moment-based structure of the estimation procedure. Further details are omitted.

\section{Standard Error Estimation} \label{sec:SE}

Asymptotic methods for standard error estimation are available for all three estimators considered in this paper. However, they rely on distinct assumptions. For the maximum likelihood and regression-based estimators, standard errors are derived conditional on the observed values of $\mathbf{X}$ and do not reflect its sampling variability. By contrast, the GMM estimator accounts for the joint distribution of $(X, Y)$ and thus incorporates uncertainty in both variables. As such, standard error estimates from the three methods are not directly comparable without adjustment, and care must be taken when interpreting their relative magnitudes.

To facilitate meaningful comparison across methods, we focus on standard error estimates that approximate the marginal variability of the estimators under a common inferential framework. In doing so, we assume that the rate parameters, $(\pi_{\mathrm{tp}}, \pi_{\mathrm{tn}})$, lie strictly in the interior of the parameter space, $(0,1)^2$, thereby excluding true boundary cases from consideration. Nonetheless, the modest sample sizes in our application (approximately 40 to 50 cases per variable), along with empirical estimates near one, raise concerns about operating in a regime where the parameters are effectively close to the boundary.

We do not invoke formal boundary asymptotics, which yield nonstandard limiting distributions \citep[see][]{self1987asymptotic}, nor do we assume failure of classical resampling methods. However, we remain mindful that proximity to the boundary (however loosely defined) can compromise the accuracy of standard asymptotic approximations. In particular, even the parametric bootstrap may perform poorly in such settings \citep[see][]{andrews2000inconsistency}. Motivated by these considerations, we explore two bootstrap procedures designed to yield more robust standard error estimates in small-sample settings. Their performance is evaluated in a simulation study in Section~\ref{sec:sim}.

The first approach is a \textit{semi-parametric bootstrap}, which treats the distribution of true counts nonparametrically while generating contaminated counts parametrically using the binomial convolution model. This hybrid approach accommodates empirical features of the true counts while preserving the underlying contamination mechanism. In contrast, a fully parametric alternative would require specifying a model for the distribution of true counts $X$, a potentially strong and unnecessary assumption in many applications. Let $\hat{\pi}_{\mathrm{tp}}$ and $\hat{\pi}_{\mathrm{tn}}$ denote the estimated true positive and true negative rates. A semi-parametric procedure bootstrap sample is then obtained as follows:
\begin{itemize}
    \item Resample the true counts $\mathbf{X}^\ast = (X_1^\ast, \ldots, X_n^\ast)$ with replacement from the empirical distribution of the observed true counts $\mathbf{X} = (X_1, \ldots, X_n)$.
    \item For each $X_i^\ast$, generate the number of true positives $\mathrm{TP}_i^\ast \sim \mathrm{Bin}(X_i^\ast, \hat{\pi}_{\mathrm{tp}})$ and false positives $\mathrm{FP}_i^\ast \sim \mathrm{Bin}(N - X_i^\ast, 1 - \hat{\pi}_{\mathrm{tn}})$.
    \item Define the contaminated count as $Y_i^\ast = \mathrm{TP}_i^\ast + \mathrm{FP}_i^\ast$, resulting in a new bootstrap sample $\mathcal{D}_{\text{semi}}^\ast = \{(X_i^\ast, Y_i^\ast),\, i=1,\ldots,n\}$.
\end{itemize}

As an alternative, we also consider the nonparametric \textit{m-out-of-n bootstrap}, which draws a subsample of size $m < n$ rather than resampling the full dataset. This approach has been shown to offer improved performance in scenarios where the standard bootstrap is inconsistent or fails, particularly in small samples or near-boundary parameter settings. The classical nonparametric bootstrap is recovered when $m = n$. In our implementation, we generate bootstrap samples by resampling $m = h(n)$ pairs $(X_i^\ast, Y_i^\ast)$ with replacement from the empirical distribution of the observed data, where $h(n)$ is a monotonic function of $n$. The resulting bootstrap sample is denoted by $\mathcal{D}_{\text{sub},m}^\ast = \{(X_i^\ast, Y_i^\ast),\, i=1,\ldots,m\}$.

In either case, the resulting bootstrap sample, whether $\mathcal{D}_{\text{semi}}^\ast$ or $\mathcal{D}_{\text{sub},m}^\ast$, is used to compute the bootstrap replicate, say $\hat{\pi}_b^\ast = \hat{\pi}(\mathcal{D}_b^\ast)$ for $b = 1, \ldots, B$, where $B$ denotes the number of bootstrap iterations. The standard deviation of the $\hat{\pi}_b^\ast$ values provides an estimate of the standard error.

For the semi-parametric bootstrap, we denote the resulting standard error by $\mathrm{SE}_{\text{semi}}^\ast$ and use it directly. For the $m$-out-of-$n$ bootstrap, we denote the unadjusted estimate by $\mathrm{SE}_{\text{sub},m}^\ast$ and apply a finite-sample correction to scale it to the original sample size $n$,
\[
\mathrm{SE}_{\text{sub}}^\ast = \left(\frac{n}{m}\right)^{1/2} \mathrm{SE}_{\text{sub},m}^\ast.
\]
This adjustment ensures that standard errors derived from the reduced bootstrap samples remain comparable to those based on the full dataset.

For the $m$-out-of-$n$ bootstrap, \citet{swanepoel1986note} recommends using $m = h(n) = \lfloor 2n/3 \rfloor$, which grows at rate $O(n)$. Other recommendations include smaller rates, such as $m = \lfloor 2\sqrt{n} \rfloor$, which is $o(n)$; see \citet{dalitz2024moonboot} for a comprehensive review of these and related considerations. While it is also possible to select $m$ adaptively based on the data, we do not pursue such approaches here.

These bootstrap methods are designed to mitigate issues that may arise when parameters lie ``near'' the boundary of the parameter space. Naturally, the semi-parametric approach presumes that the assumed generative model is correctly specified, whereas the $m$-out-of-$n$ approach trades off some efficiency for increased robustness. As noted earlier, we evaluate the performance of both methods in the simulation study, focusing on their behavior under small-sample conditions and near-boundary scenarios.

\section{Simulation Study} \label{sec:sim}

To evaluate the performance of the proposed estimation methods, we conducted three simulation studies, each designed to assess a distinct aspect of estimator behavior. The first study examined accuracy under correct model specification. Specifically, we focus on how sample size, misclassification rates, and dispersion influence root mean square error (RMSE) when the binomial convolution model is the true data-generating mechanism for contaminated counts. The second study assessed robustness to model misspecification, where the binomial convolution assumption is violated. The third study evaluated the performance of different standard error estimation strategies, including asymptotic and bootstrap-based approaches as described in Section~\ref{sec:SE}, again under the correctly specified model.

\subsection{Estimator Accuracy Under the Binomial Convolution Model} \label{sec:sim_estim_accuracy}

We begin by evaluating the finite-sample performance of each method in recovering the true positive rate $\pi_{\mathrm{tp}}$ and true negative rate $\pi_{\mathrm{tn}}$ under the correctly specified binomial convolution model. A baseline data-generating mechanism was defined with sample size $n = 50$, number of trials $N = 60$, true classification rates $\pi_{\mathrm{tp}} = 0.98$ and $\pi_{\mathrm{tn}} = 0.70$, and a true success probability $p = 0.95$. In each replication, independent true counts were generated as $X_i \sim \text{Binomial}(N, p)$ for $i = 1, \dots, n$, and contaminated counts $Y_i$ were then generated according to the binomial convolution model described in Section~\ref{sec:Model_form}.

To evaluate sensitivity to key model parameters, we varied one factor at a time while holding the others fixed. Specifically, we varied the sample size $n$ from 30 to 100 (in steps of 5); generated 15 equally spaced values of $\pi_{\mathrm{tp}} \in [0.85, 0.999]$ and $\pi_{\mathrm{tn}} \in [0.5, 0.95]$; and introduced overdispersion in the true counts by generating $X_i$ from a beta-binomial distribution with intra-class correlation $\rho \in [0, 0.06]$, also in 15 equally spaced steps. These parameter ranges were chosen to reflect the empirical setting described in Section~\ref{sec:data_application}. While $\rho = 0$ corresponds to the binomial case, the upper bound $\rho = 0.06$ represents substantial overdispersion. Specifically, this implies a multiplicative variance inflation factor of $\nu = 1 + (n - 1)\rho = 1 + 49 \cdot 0.06 \approx 4$, yielding a variance nearly four times larger than the binomial variance $Np(1-p)$.

\begin{figure}[]
\centering
\includegraphics[width=0.8\textwidth]{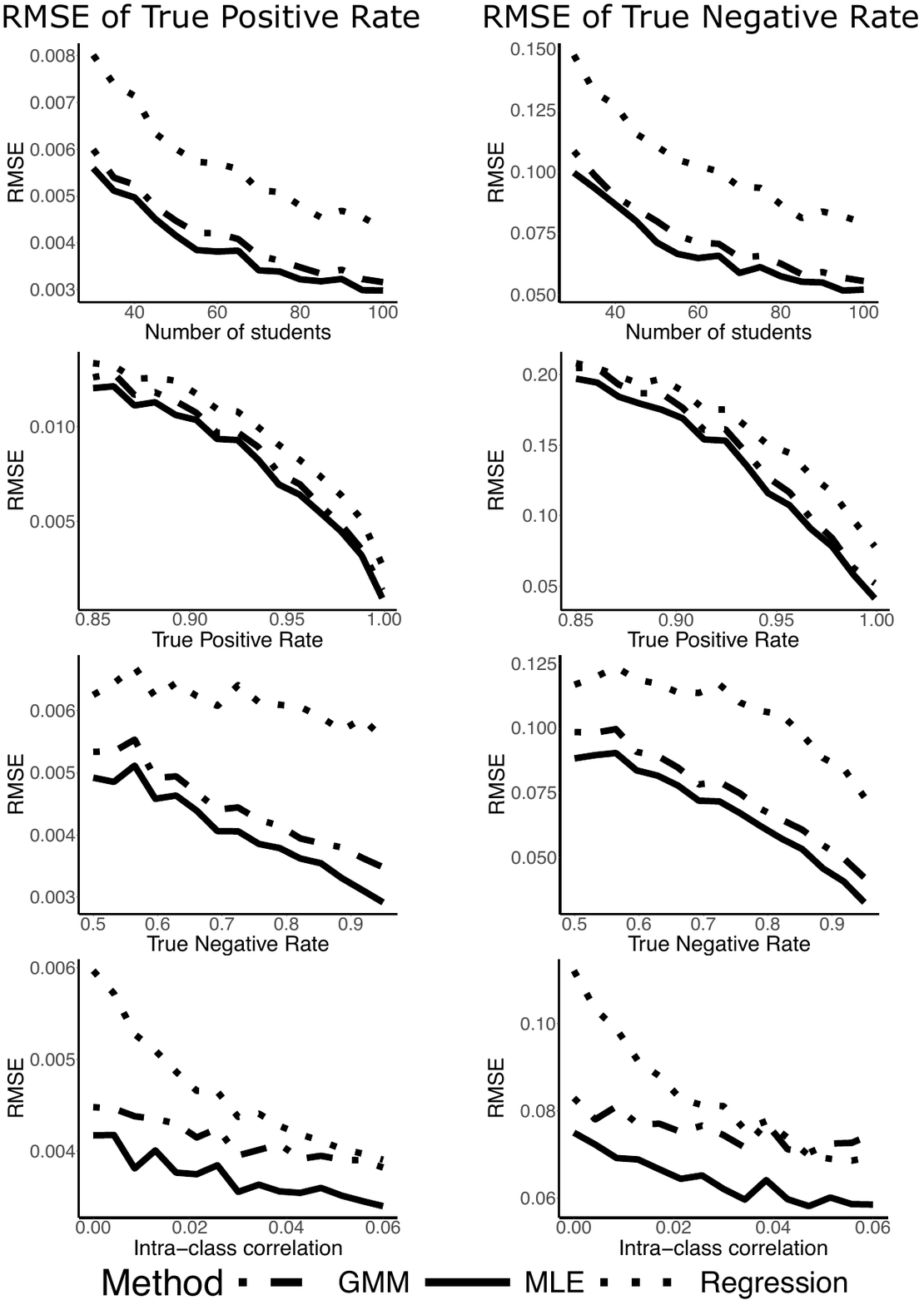}
\caption{RMSE of $\hat{\pi}_{\text{tp}}$ and $\hat{\pi}_{\text{tn}}$ under the correctly specified binomial convolution model, shown across varying sample sizes, misclassification rates, and intra-class correlations.}
\label{rmse_comparison}
\end{figure}

This design yielded 60 distinct simulation conditions. For each, we generated 1{,}000 datasets and applied all three estimation methods (MLE, regression, and GMM) to recover $\pi_{\mathrm{tp}}$ and $\pi_{\mathrm{tn}}$. Root mean square error (RMSE) was used as the primary performance metric and is summarized in Figure~\ref{rmse_comparison}.

Across all simulation conditions, a consistent ranking of estimator performance emerged: MLE achieved the lowest RMSE, followed by GMM, with the regression approach performing worst. This ordering reflects the amount of model information each method leverages; MLE uses the full likelihood, GMM relies on first- and second-order moments, and regression depends only on the first conditional moment. In general, the regression estimator exhibited substantially higher RMSE than the other two methods.

When considering the influence of the specific parameters varied, several noteworthy patterns emerge. Estimation accuracy improved with increasing sample size and with higher values of $\pi_{\mathrm{tp}}$ and $\pi_{\mathrm{tn}}$, which make the contamination process more deterministic by introducing fewer classification errors. RMSE also declined with increasing over-dispersion in the true counts (via larger values of intra-class correlation $\rho$). This improvement is attributed to greater variability in $X$, which results in a wider range of observed true count values and, in turn, better identification of model parameters. This phenomenon parallels the role of increased design variance in linear regression, where broader spread in predictor values yields more precise estimates.

While GMM consistently outperformed regression overall, the gap between them narrowed substantially at higher levels of over-dispersion. For small values of $\rho$, regression showed markedly higher RMSE than GMM, but the two methods yielded comparable performance as $\rho$ increased. This convergence, evident in Figure~\ref{rmse_comparison}, suggests that greater heterogeneity in $X$ may help mitigate regression’s limitations relative to GMM.

\FloatBarrier

\subsection{Performance Under Model Misspecification} \label{sec:sim_misspec}

We next assessed the robustness of each estimator to model misspecification by introducing overdispersion into the contamination process. Specifically, the conditional binomial components used to generate true positive and true negative counts were replaced with beta-binomial distributions, thereby violating the assumptions of the binomial convolution model under which estimation was still carried out. Three misspecification scenarios were considered: (1) overdispersion applied only to the true positive counts, (2) only to the true negative counts, and (3) to both. This design allows us to examine plausible ways in which the binomial convolution model may be violated while maintaining control over the source of misspecification.

All other aspects of the data-generating process were fixed to the baseline values described in Section~\ref{sec:sim_estim_accuracy}, with $n = 50$, $N = 60$, $p = 0.95$, $\pi_{\mathrm{tp}} = 0.98$, and $\pi_{\mathrm{tn}} = 0.70$. True counts $X_i$ were generated independently as $\mathrm{Binomial}(N, p)$. Model misspecification was introduced by replacing the binomial components of the contamination process with beta-binomial distributions. Specifically, when overdispersion was introduced in the true positive component, we generated $\mathrm{TP}_i \sim \mathrm{BetaBin}(X_i, \pi_{\mathrm{tp}}, \rho)$; when overdispersion was introduced in the true negative component, we generated $\mathrm{FP} \sim \mathrm{BetaBin}(N - X_i, 1 - \pi_{\mathrm{tn}}, \rho)$. The intra-class correlation $\rho$ was varied over 15 equally spaced values in the interval $[0, 0.06]$. For each combination of contamination scenario and $\rho$, we generated 1{,}000 datasets and applied all three estimators to recover $\pi_{\mathrm{tp}}$ and $\pi_{\mathrm{tn}}$.

\begin{sidewaysfigure}
\centering
\includegraphics[width=0.45\textwidth]{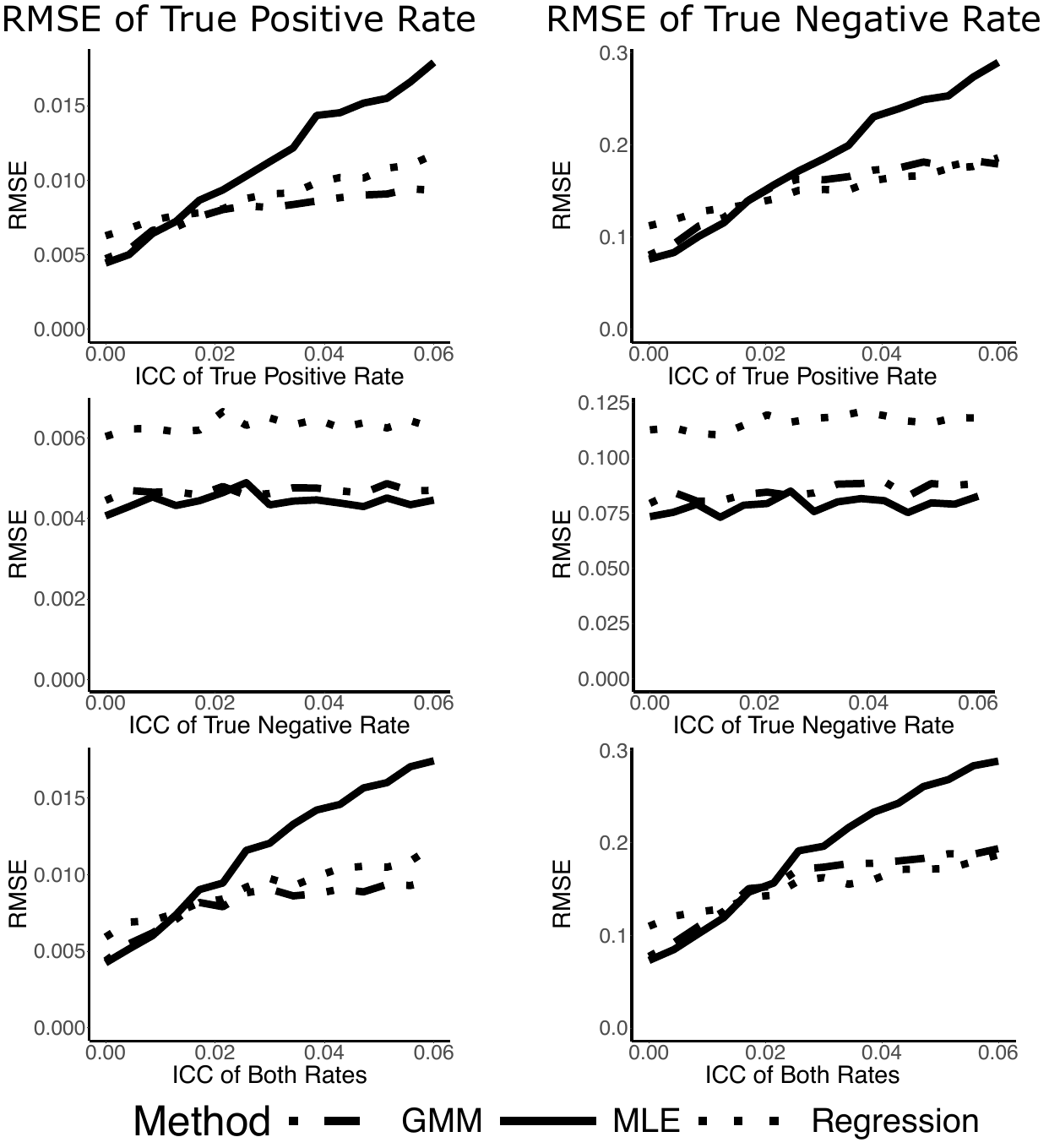}\hfill
\includegraphics[width=0.45\textwidth]{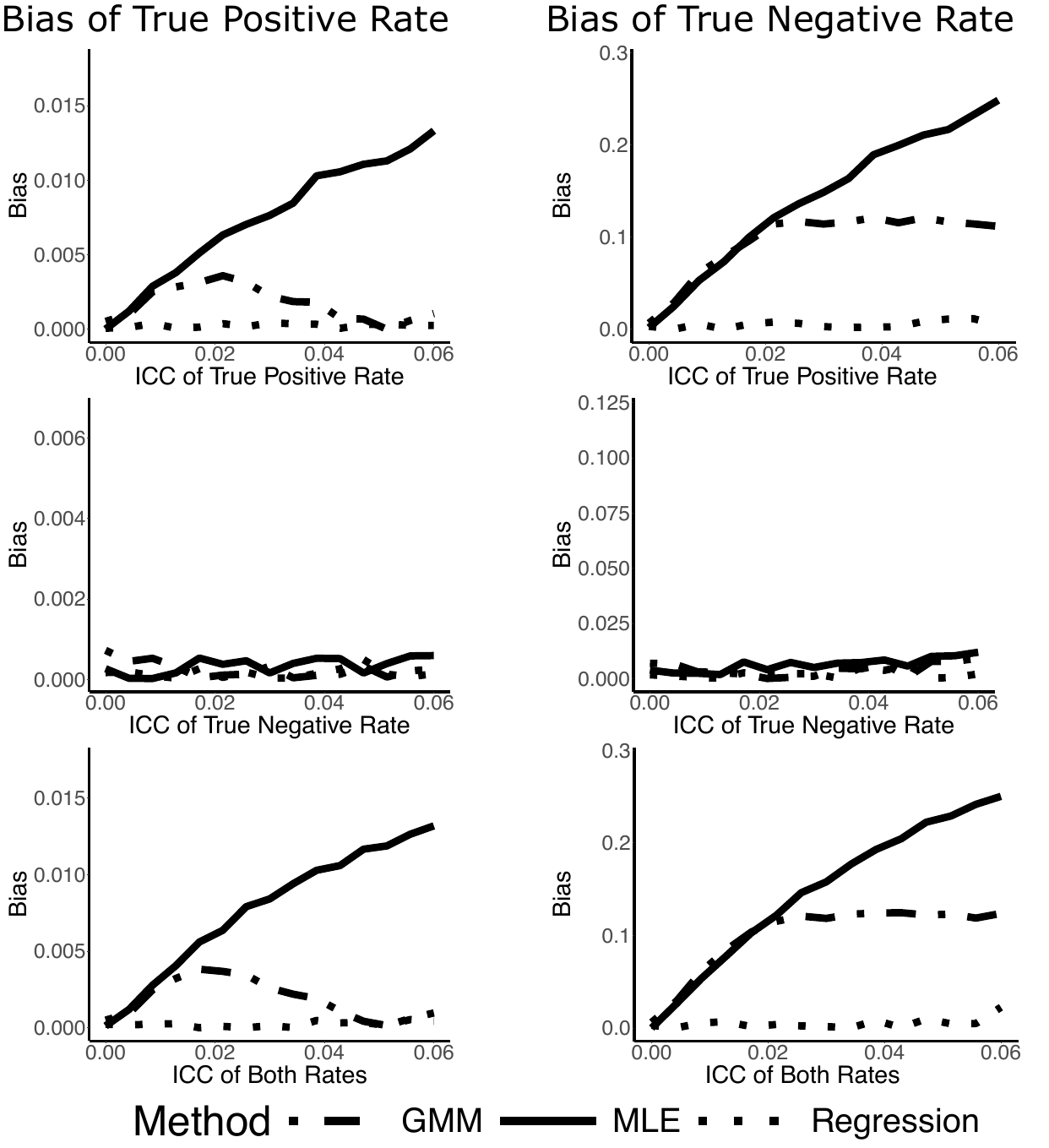}
\caption{RMSE and bias under increasing levels of model misspecification.}
\label{model_misspec}
\end{sidewaysfigure}

Figure~\ref{model_misspec} presents the RMSE and absolute bias for each estimator across these settings. When overdispersion was introduced into the true positive component (either alone or in conjunction with the true negative component), the MLE deteriorated rapidly with increasing $\rho$, largely due to increasing bias. This pattern reflects the estimator’s sensitivity to violations of its fully specified likelihood assumptions. In contrast, both the regression and GMM estimators were notably more stable under misspecification. The regression estimator, in particular, exhibited the lowest bias across most scenarios, although its variance remained comparatively high.

When overdispersion was applied only to the true negative counts, however, MLE remained the best-performing method in terms of RMSE. This likely reflects the limited influence of the true negative component in the simulated data. Specifically, with a high true success probability $p$ and moderate number of trials $N$, the resulting true counts $X_i$ are typically close to $N$, leaving few failures ($N - X_i$) from which false positives can arise. As a result, even when the false positive counts $FP_i$ are drawn from an overdispersed distribution, the overall contribution of this component to estimation error remains small. The relative insensitivity of all three methods to overdispersion in the true negative channel further supports this interpretation. A supplementary simulation (not reported here) with $p = 0.5$ confirms the increased sensitivity of MLE to overdispersion in the true negative counts when false positives occur more frequently.

These results underscore important trade-offs. While MLE is most efficient under correct specification, it is highly vulnerable to violations in the contamination mechanism. Conversely, the regression and GMM estimators offer improved robustness in such cases—particularly when misspecification affects the more consequential true positive channel. For clarity, we use “robustness” here to refer to reduced sensitivity to contamination model misspecification, as measured by the resulting bias and RMSE.

\subsection{Evaluation of Standard Error Estimation Methods} \label{sec:sim_se}

Finally, we assessed the accuracy of standard error estimation under a correctly specified model. A fully crossed simulation design was used with two levels each for five factors: number of trials $N \in {44, 69}$, true positive rate $\pi_{\mathrm{tp}} \in {0.96, 0.999}$, true negative rate $\pi_{\mathrm{tn}} \in {0.75, 0.85}$, true success probability $p \in {0.96, 0.98}$, and intra-class correlation $\rho \in {0, 0.03}$. This yielded 32 unique simulation settings. For each setting, we generated 1{,}000 datasets and computed point estimates using all three estimation methods. Standard errors were then estimated using method-specific plug-in formulas (asymptotic estimators), the semi-parametric bootstrap, the classic nonparametric bootstrap, and two versions of the $m$-out-of-$n$ bootstrap with $m = \lfloor 2\sqrt{n} \rfloor$ and $m = \lfloor 2n/3 \rfloor$, respectively. All bootstrap applications used $B = 50$ replicates, which we found to offer a reasonable tradeoff between computational burden and stability of the results.

We evaluated the accuracy of each variance estimation method by comparing the estimated variance to the empirical Monte Carlo variance of the point estimates across the 1{,}000 replications per scenario. These comparisons are summarized as variance ratios, defined as
$\mathrm{Ratio}_{\text{method}} = {\mathrm{AV}}_{\text{method}} / \mathrm{EV}_{\text{mc}},$
where ${\mathrm{AV}}_{\text{method}}$ denotes the \textit{average} of the variance estimates produced by the method across the 1{,}000 simulated datasets, and $\mathrm{EV}_{\text{mc}}$ is the empirical variance of the point estimates over those same replications. A variance ratio centered at 1 indicates approximate unbiasedness; full results for each setting are provided in Appendix~\ref{sec:var_ratio_tables}.

Results varied notably by estimator. For the regression estimator, the plug-in approach was approximately unbiased on average, as indicated by median variance ratios near 1, but it exhibited high variability across scenarios. The semi-parametric bootstrap performed well, yielding slight upward bias in variance estimates. Both the classic and $m$-out-of-$n$ bootstrap methods tended to underestimate the variance.

In contrast, variance estimation for GMM proved more difficult. The plug-in estimator was particularly poor, exhibiting substantial downward bias. Bootstrap-based approaches performed comparatively better for GMM (partly in the sense that overestimation is conservatively preferable to underestimation) with variance ratios typically centered between 1.15 and 1.25. All variance estimation methods here had high variability.

For the MLE, the plug-in method slightly overestimated the variance, while all bootstrap approaches yielded accurate and consistent results, with variance ratios tightly concentrated around 1. Overall, MLE variance estimates were the most stable (median ratios close to 1 and narrow dispersion), GMM the least, and regression fell between the two.

\begin{figure}[h!]
\centering
\includegraphics[width=\textwidth]{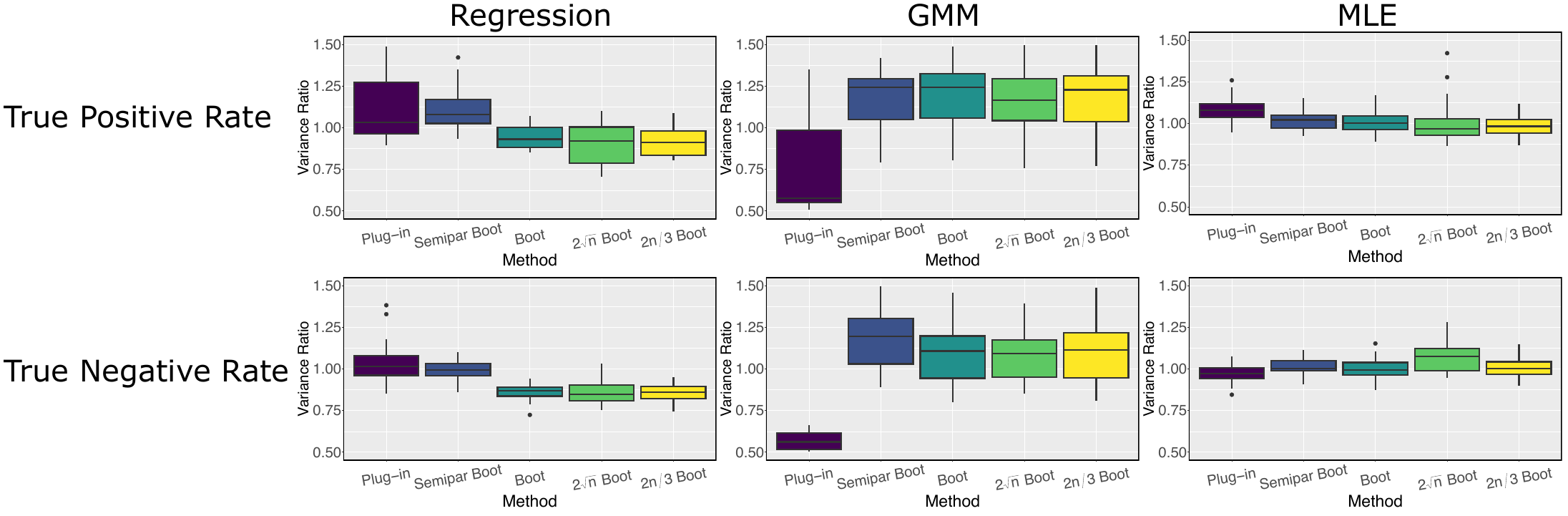}
\caption{Variance ratio boxplots for simulation scenarios}
\label{variance_ratio}
\end{figure}

These findings underscore the importance of aligning standard error estimation methods with the structural characteristics of the estimator. In particular, GMM remains challenging in the small-sample settings considered and future work should consider refinement to yield reliable uncertainty estimates. When the binomial convolution model is deemed correctly specified, the semi-parametric bootstrap is preferred. The $m$-out-of-$n$ bootstrap offered no clear advantage over the classical bootstrap in the context of this study, despite choosing true parameter values ``close'' to the boundary of the parameter space relative to the sample size.

\section{Real Data Application} \label{sec:data_application}

We now turn to the motivating application: an oral reading fluency (ORF) assessment dataset drawn from a substudy of the CORE project \citep{nese2014measuring}, comprising student responses to 10 distinct reading passages. Each student read a single passage, yielding a set of independent observations denoted $(x_i, y_{1i}, y_{2i})$ for $i = 1, \ldots, n$. Here, $x_i$ is the true number of words read correctly, obtained via multi-rater consensus and assumed to be error-free; $y_{1i}$ is the score assigned by a single human rater under simulated classroom conditions; and $y_{2i}$ is the automated score (hereafter referred to as AI scoring), generated by specialized voice recognition software that aligns the recorded reading with the target passage text to estimate the number of words read correctly. Each observation is associated with a passage index $d(i) := d \in {1, \ldots, 10}$, where the corresponding passage length is denoted $N_d$, and the number of students assessed on passage $d$ is $n_d$, such that $\sum_d n_d = n$. Passage lengths $N_d$ range from 44 to 69 words, with $n_d$ ranging from 42 to 51 students per passage.

\begin{figure}[h]
\centering
\includegraphics[width=0.9\textwidth]{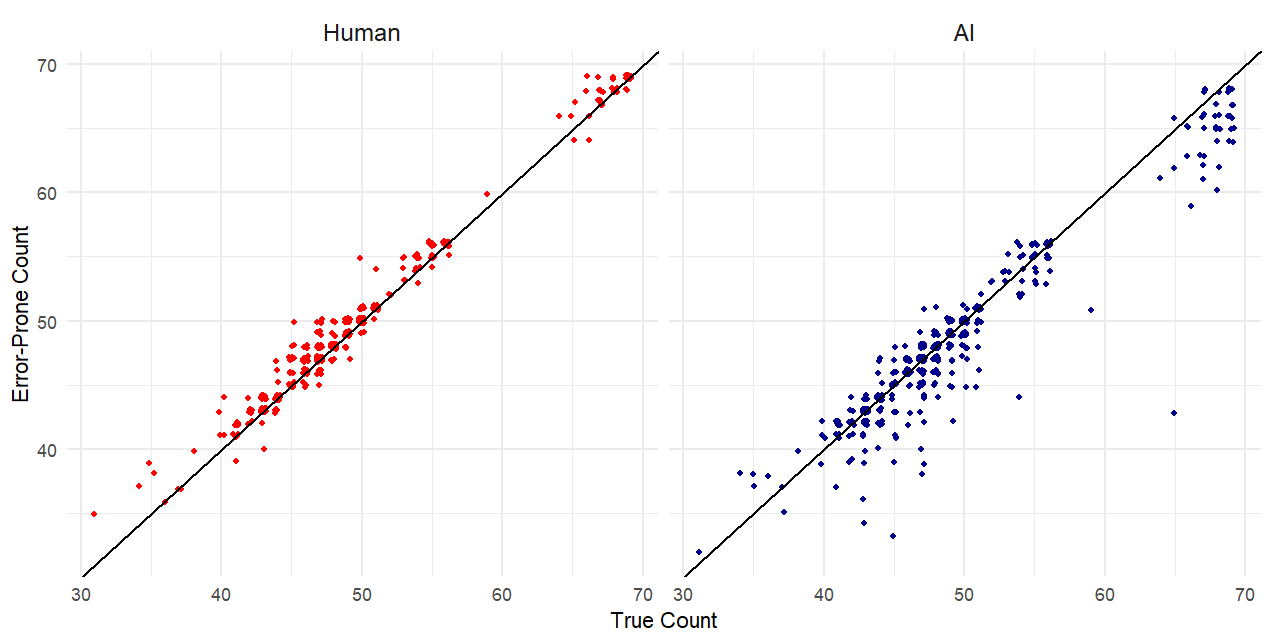}
\caption{Scatterplots of error-prone word reading counts versus true counts across all 10 passages.}
\label{fig:true_vs_error_counts}
\end{figure}

Figure~\ref{fig:true_vs_error_counts} presents scatterplots of observed (error-prone) word counts versus true scores, with separate panels for human and AI scoring. To reduce overplotting due to discreteness, points are jittered. Each point corresponds to a single student’s reading of a single passage. The 45-degree reference line indicates perfect agreement with the true score; deviations from this line reflect the magnitude and direction of measurement error. Human-assigned scores tend to slightly overestimate the true count, while AI-generated scores display greater variability and are more likely to underestimate.

To further illustrate the discrepancies between the two error-prone scoring methods, Table~\ref{tab:accuracy_props} summarizes the proportion of observations that exactly match the true count, differ by one point, or differ by more than one point. Human scores matched the true count in 64.3\% of cases, compared to 42.2\% for AI-generated scores. Larger deviations were notably more common for AI scoring, with 25.4\% of cases differing by more than one point, versus just 9.3\% for human ratings. These patterns highlight the importance of explicitly modeling measurement error in subsequent analyses.

\begin{table}[h]
\small
\centering
\caption{Proportion of scores matching or deviating from the true count, by rater type}
\vspace{0.5em}
\label{tab:accuracy_props}
\begin{tabular}{lcc}
\toprule
 & {Human} & {AI} \\
\midrule
Exact Match (True Score) & 0.643 & 0.422 \\
One Point Off              & 0.264 & 0.324 \\
More Than Point Off           & 0.093 & 0.254 \\
\bottomrule
\end{tabular}
\end{table}

\begin{figure}[h!]
\centering
\includegraphics[width=\textwidth]{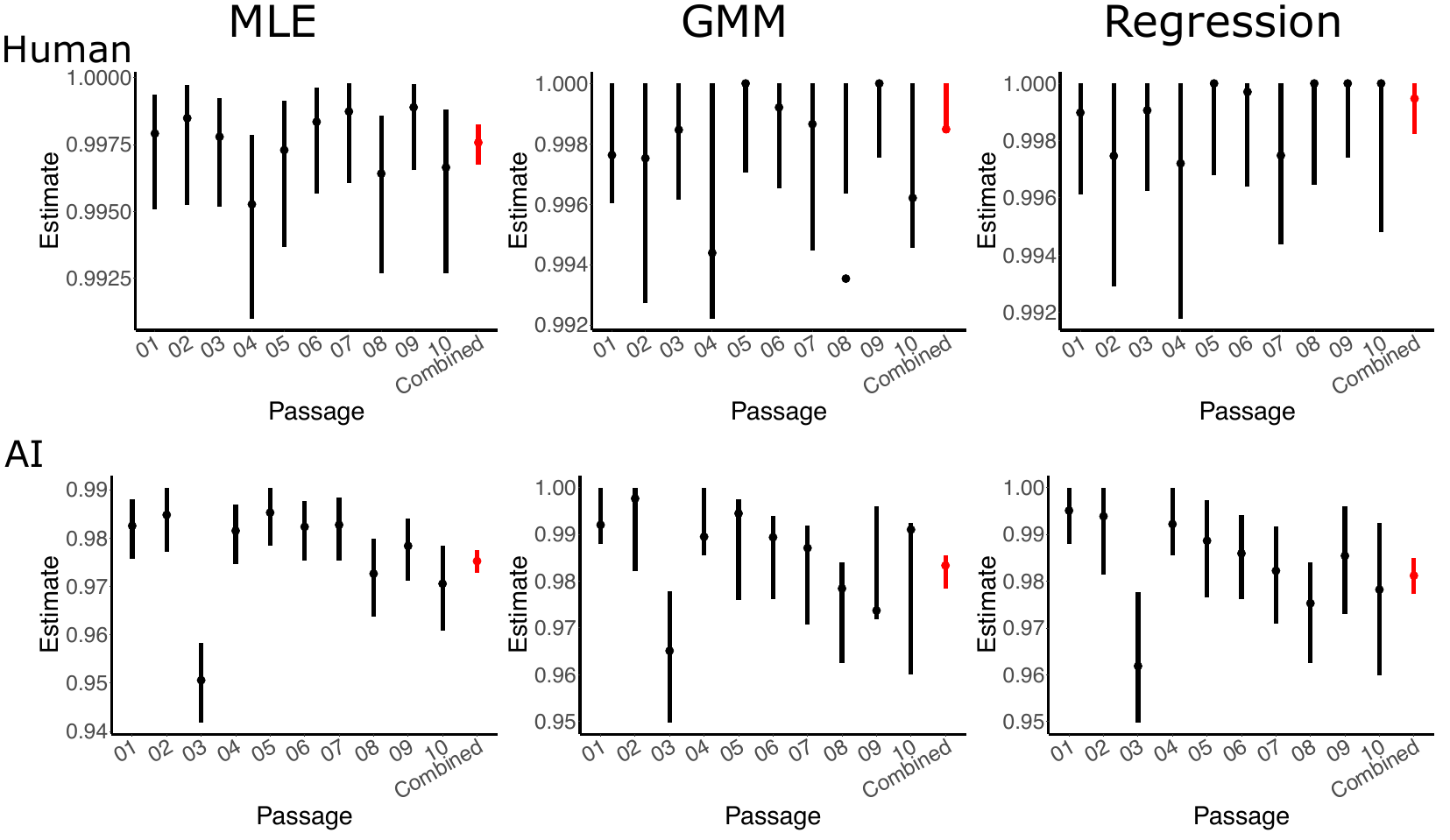}
\caption{Estimated true positive rates for human and AI scoring across 10 reading passages. Results are shown for each estimation method, with 95\% confidence intervals.}

\label{real_tp}
\end{figure}

\begin{figure}[h!]
\centering
\includegraphics[width=\textwidth]{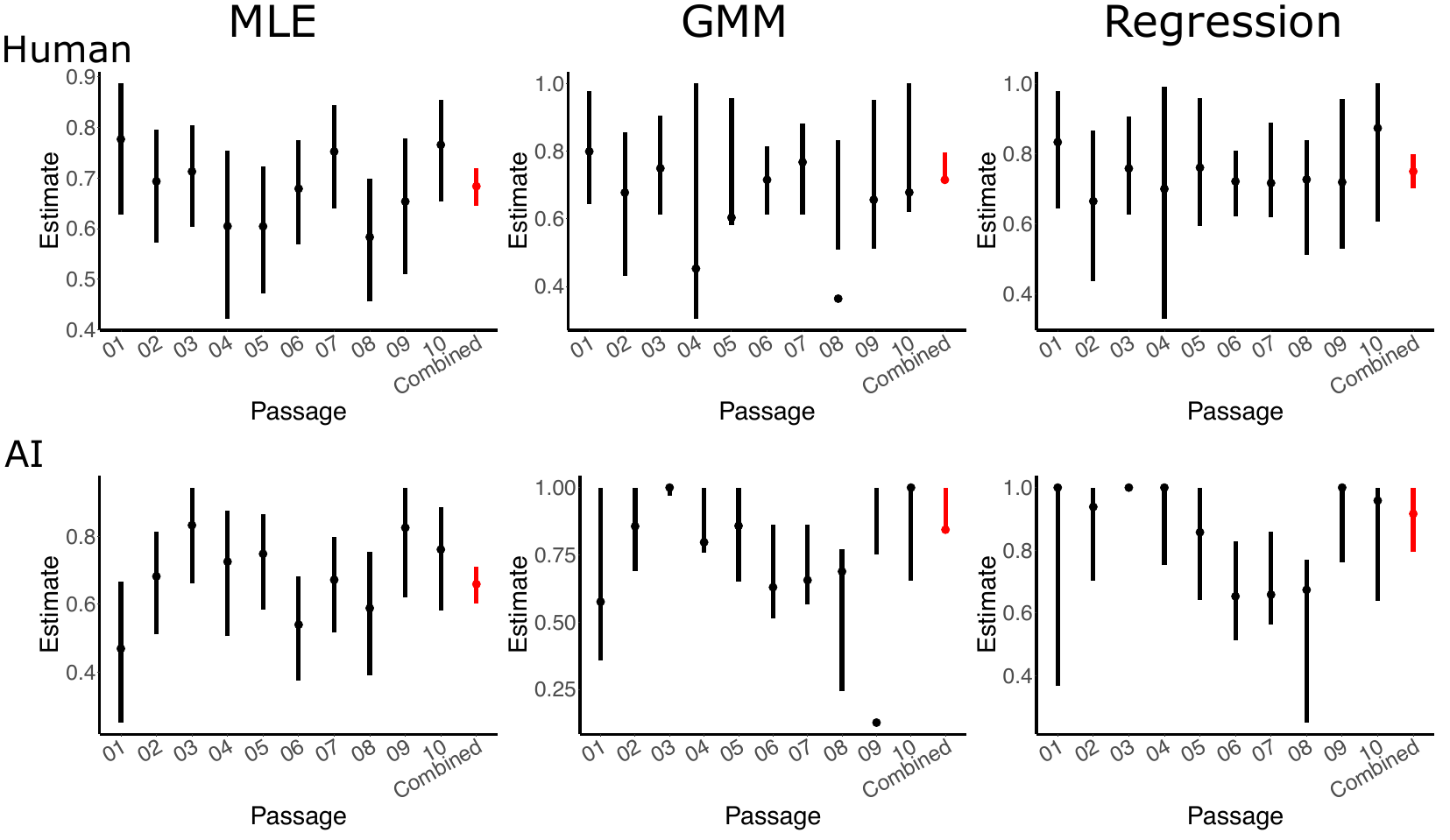}
\caption{Estimated true negative rates for human and AI scoring across 10 reading passages. Results are shown for each estimation method, with 95\% confidence intervals.}

\label{real_tn}
\end{figure}

For each passage, we applied all three estimation methods to obtain point estimates of the true positive and true negative rates for both human- and AI-assigned scores. In addition to these passage-specific estimates, where parameters are estimated separately for each passage, we also report combined estimates under the assumption that all passages share common parameters. Point estimates and associated 95\% confidence intervals are presented in Figures~\ref{real_tp} and~\ref{real_tn}, while the corresponding point estimates and standard errors are reported numerically in Tables~\ref{tab:human_parms} and~\ref{tab:auto_parms}. Note that two observations from passage 8 and one from passage 9 were excluded from the analysis due to the GMM estimator’s sensitivity to outliers. To ensure consistency and comparability across methods, these outliers were excluded from all analyses. Their specific impact on GMM, which was minimal for both MLE and regression, is discussed in detail below.

For the maximum likelihood estimates, 95\% confidence intervals were constructed using the likelihood ratio method, with standard errors derived from the observed information matrix. For the regression and GMM approaches, confidence intervals were computed from the empirical quantiles of 2{,}000 bootstrap replicates, and standard errors were similarly obtained via the standard nonparametric bootstrap. These choices were informed by the simulation results in Section~\ref{sec:sim_se}. As demonstrated there, standard error estimation for GMM is especially sensitive to small-sample variability. Accordingly, the GMM confidence intervals and standard errors reported here should be interpreted with caution, as they may be biased or exhibit inflated variability.

\FloatBarrier

\begin{table}
\small
\caption{Point estimates and corresponding standard errors for human scoring accuracy parameters, reported by estimation method and passage.}
\label{tab:human_parms}
\centering
\begin{tabular}[t]{lccccccp{15mm}}
\toprule
  & \multicolumn{3}{c}{True Positive $(\hat{\pi}_{\mathrm{tp}})$} & \multicolumn{3}{c}{True Negative $(\hat{\pi}_{\mathrm{tn}})$} \\
  \cmidrule(lr){2-4} \cmidrule(lr){5-7}
  & MLE & GMM & LS & MLE & GMM & LS\\
\midrule
Passage 1 & \shortstack{0.9979 \\ (0.0011)} &  \shortstack{0.9976 \\ (0.0009)} &  \shortstack{0.9990 \\ (0.0012)} &  \shortstack{0.7767 \\ (0.0663)} &  \shortstack{0.7990 \\ (0.0734)} &  \shortstack{0.8328 \\ (0.0932)}\\
\addlinespace
Passage 2 & \shortstack{0.9985 \\ (0.0011)} &  \shortstack{0.9975 \\ (0.0016)} &  \shortstack{0.9975 \\ (0.0021)} &  \shortstack{0.6934 \\ (0.0570)} &  \shortstack{0.6775 \\ (0.0926)} &  \shortstack{0.6648 \\ (0.1187)}\\
\addlinespace
Passage 3 & \shortstack{0.9978 \\ (0.0010)} &  \shortstack{0.9985 \\ (0.0009)} &  \shortstack{0.9991 \\ (0.0012)} &  \shortstack{0.7129 \\ (0.0512)} &  \shortstack{0.7492 \\ (0.0573)} &  \shortstack{0.7579 \\ (0.0665)}\\
\addlinespace
Passage 4 & \shortstack{0.9953 \\ (0.0017)} &  \shortstack{0.9944 \\ (0.0017)} &  \shortstack{0.9972 \\ (0.0023)} &  \shortstack{0.6048 \\ (0.0844)} &  \shortstack{0.4530 \\ (0.1465)} &  \shortstack{0.6998 \\ (0.1794)}\\
\addlinespace
Passage 5 & \shortstack{0.9973 \\ (0.0014)} &  \shortstack{1.0000 \\ (0.0006)} &  \shortstack{1.0000 \\ (0.0010)} &  \shortstack{0.6045 \\ (0.0637)} &  \shortstack{0.6036 \\ (0.0831)} &  \shortstack{0.7603 \\ (0.1037)}\\
\addlinespace
Passage 6 & \shortstack{0.9983 \\ (0.0010)} &  \shortstack{0.9992 \\ (0.0008)} &  \shortstack{0.9997 \\ (0.0011)} &  \shortstack{0.6789 \\ (0.0523)} &  \shortstack{0.7151 \\ (0.0423)} &  \shortstack{0.7211 \\ (0.0508)}\\
\addlinespace
Passage 7 & \shortstack{0.9987 \\ (0.0009)} &  \shortstack{0.9987 \\ (0.0012)} &  \shortstack{0.9975 \\ (0.0016)} &  \shortstack{0.7523 \\ (0.0522)} &  \shortstack{0.7675 \\ (0.0583)} &  \shortstack{0.7164 \\ (0.0734)}\\
\addlinespace
Passage 8 & \shortstack{0.9964 \\ (0.0015)} &  \shortstack{0.9987 \\ (0.0008)} &  \shortstack{1.0000 \\ (0.0012)} &  \shortstack{0.5792 \\ (0.0660)} &  \shortstack{0.7545 \\ (0.0854)} &  \shortstack{0.7511 \\ (0.1152)}\\
\addlinespace
Passage 9 & \shortstack{0.9989 \\ (0.0008)} &  \shortstack{1.0000 \\ (0.0006)} &  \shortstack{1.0000 \\ (0.0008)} &  \shortstack{0.6536 \\ (0.0682)} &  \shortstack{0.6562 \\ (0.0945)} &  \shortstack{0.7189 \\ (0.1200)}\\
\addlinespace
Passage 10 & \shortstack{0.9966 \\ (0.0016)} &  \shortstack{0.9962 \\ (0.0013)} &  \shortstack{1.0000 \\ (0.0016)} &  \shortstack{0.7656 \\ (0.0515)} &  \shortstack{0.6780 \\ (0.0908)} &  \shortstack{0.8727 \\ (0.1184)}\\
\addlinespace
Combined & \shortstack{0.9976 \\ (0.0004)} &  \shortstack{0.9985 \\ (0.0004)} &  \shortstack{0.9995 \\ (0.0005)} &  \shortstack{0.6847 \\ (0.0192)} &  \shortstack{0.7173 \\ (0.0206)} &  \shortstack{0.7529 \\ (0.0273)}\\
\bottomrule
\end{tabular}
\end{table}

\begin{table}
\small
\caption{Point estimates and corresponding standard errors for AI scoring accuracy parameters, reported by estimation method and passage.}
\label{tab:auto_parms}
\centering
\begin{tabular}[t]{lcccccc}
\toprule
  & \multicolumn{3}{c}{True Positive $(\hat{\pi}_{\mathrm{tp}})$} & \multicolumn{3}{c}{True Negative $(\hat{\pi}_{\mathrm{tn}})$} \\
\cmidrule(lr){2-4} \cmidrule(lr){5-7}
  & MLE & GMM & LS & MLE & GMM & LS\\
\midrule
Passage 1 & \shortstack{0.9825 \\ (0.0031)} &  \shortstack{0.9920 \\ (0.0028)} &  \shortstack{0.9951 \\ (0.0036)} &  \shortstack{0.4692 \\ (0.1062)} &  \shortstack{0.5757 \\ (0.1672)} &  \shortstack{1.0000 \\ (0.2052)}\\
\addlinespace
Passage 2 & \shortstack{0.9848 \\ (0.0034)} &  \shortstack{0.9977 \\ (0.0042)} &  \shortstack{0.9939 \\ (0.0054)} &  \shortstack{0.6823 \\ (0.0774)} &  \shortstack{0.8561 \\ (0.0803)} &  \shortstack{0.9384 \\ (0.0987)}\\
\addlinespace
Passage 3 & \shortstack{0.9506 \\ (0.0042)} &  \shortstack{0.9651 \\ (0.0056)} &  \shortstack{0.9619 \\ (0.0075)} &  \shortstack{0.8331 \\ (0.0716)} &  \shortstack{1.0000 \\ (0.0366)} &  \shortstack{1.0000 \\ (0.0478)}\\
\addlinespace
Passage 4 & \shortstack{0.9815 \\ (0.0031)} &  \shortstack{0.9895 \\ (0.0030)} &  \shortstack{0.9922 \\ (0.0039)} &  \shortstack{0.7257 \\ (0.0947)} &  \shortstack{0.7966 \\ (0.0591)} &  \shortstack{1.0000 \\ (0.0734)}\\
\addlinespace
Passage 5 & \shortstack{0.9853 \\ (0.0030)} &  \shortstack{0.9945 \\ (0.0043)} &  \shortstack{0.9887 \\ (0.0054)} &  \shortstack{0.7487 \\ (0.0716)} &  \shortstack{0.8578 \\ (0.0748)} &  \shortstack{0.8574 \\ (0.0932)}\\
\addlinespace
Passage 6 & \shortstack{0.9823 \\ (0.0031)} &  \shortstack{0.9894 \\ (0.0036)} &  \shortstack{0.9859 \\ (0.0047)} &  \shortstack{0.5397 \\ (0.0782)} &  \shortstack{0.6295 \\ (0.0753)} &  \shortstack{0.6529 \\ (0.0884)}\\
\addlinespace
Passage 7 & \shortstack{0.9827 \\ (0.0033)} &  \shortstack{0.9871 \\ (0.0042)} &  \shortstack{0.9822 \\ (0.0055)} &  \shortstack{0.6720 \\ (0.0721)} &  \shortstack{0.6558 \\ (0.0634)} &  \shortstack{0.6582 \\ (0.0813)}\\
\addlinespace
Passage 8 & \shortstack{0.9726 \\ (0.0041)} &  \shortstack{0.9784 \\ (0.0044)} &  \shortstack{0.9753 \\ (0.0057)} &  \shortstack{0.5885 \\ (0.0928)} &  \shortstack{0.6883 \\ (0.1167)} &  \shortstack{0.6734 \\ (0.1454)}\\
\addlinespace
Passage 9 & \shortstack{0.9795 \\ (0.0033)} &  \shortstack{0.9928 \\ (0.0048)} &  \shortstack{0.9831 \\ (0.0062)} &  \shortstack{0.7935 \\ (0.0977)} &  \shortstack{1.0000 \\ (0.0934)} &  \shortstack{0.9825 \\ (0.1145)}\\
\addlinespace
Passage 10 & \shortstack{0.9705 \\ (0.0045)} &  \shortstack{0.9910 \\ (0.0067)} &  \shortstack{0.9782 \\ (0.0087)} &  \shortstack{0.7615 \\ (0.0779)} &  \shortstack{1.0000 \\ (0.0855)} &  \shortstack{0.9585 \\ (0.1110)}\\
\addlinespace
Combined & \shortstack{0.9753 \\ (0.0012)} &  \shortstack{0.9832 \\ (0.0015)} &  \shortstack{0.9811 \\ (0.0021)} &  \shortstack{0.6560 \\ (0.0280)} &  \shortstack{0.8350 \\ (0.0444)} &  \shortstack{0.9081 \\ (0.0635)}\\
\bottomrule
\end{tabular}
\end{table}

Regarding the outliers noted earlier, their inclusion led to substantial shifts in GMM-estimated parameters: for passage 8, the human-scored true negative rate $\hat{\pi}_{\mathrm{tn}}$ dropped to 0.364, compared to 0.755 when the outliers were excluded; for passage 9, the AI-scored true negative rate declined sharply to 0.126, versus 1.000 with the outlier removed. Removing these outliers brought the GMM estimates into closer alignment with those obtained from MLE and regression, both of which were minimally affected by the outliers. This underscores the relative robustness of MLE and regression methods in the presence of influential observations.

To conclude the data illustration, we note that in this particular instance, human raters exhibited higher overall accuracy and, notably, greater consistency across passages than the AI scoring system. Model selection using AIC and BIC supports this conclusion. When comparing a common-parameter model (assuming constant accuracy rates across passages) to a passage-specific model, the simpler specification was strongly preferred for human ratings (AIC = 849.2, BIC = 857.5 vs. AIC = 866.7, BIC = 950.3), suggesting that human scoring accuracy does not meaningfully vary by passage. In contrast, for AI scoring, the passage-specific model was favored (AIC = 1830.5, BIC = 1914.1 vs. AIC = 1923.1, BIC = 1931.4), indicating greater heterogeneity in AI performance. This variability is also reflected in Figures~\ref{real_tp} and~\ref{real_tn}, where AI scoring estimates can be seen to vary by passage. While further analysis is needed to determine whether this reflects systematic passage-level effects or isolated anomalies, these results suggest that consistency in automated scoring warrants continued scrutiny. Importantly, as speech recognition and natural language processing technologies continue to evolve, such limitations may not persist, and future iterations of automated scoring systems may yield improved accuracy and consistency.

\section{Discussion}

Measurement error in count data remains an underexplored area, particularly when errors resemble misclassification but arise from aggregated event counts. This is illustrated in our motivating example, where the number of successes in a fixed number of trials is recorded, but the recorded count is subject to error. This paper introduces a novel binomial convolution model that formally extends ideas from binary misclassification to bounded count settings while preserving the discrete and bounded nature of the data. By incorporating distinct true positive and true negative rates, it captures asymmetric measurement error in a manner that reflects realistic data-generating processes in applied contexts.

We proposed and compared three estimation strategies under this model: maximum likelihood estimation (MLE), linear regression, and a generalized method of moments (GMM) estimator. Each method strikes a different balance among model fidelity, statistical robustness, and computational complexity. Extensive simulations show that MLE performs best when the binomial convolution model is correctly specified, though it is computationally intensive due to its reliance on convolution-based likelihoods and can be sensitive to model misspecification. Regression, by contrast, is simple to implement and interpret, but typically exhibits higher root mean square error. The GMM estimator, which relies only on the first two moments of the model, proved unexpectedly sensitive to outliers in our data application. For this reason, we currently recommend MLE and regression as more reliable estimation methods.

The oral reading fluency (ORF) application highlights the practical challenges associated with real-world estimation. Here, the inclusion of three influential observations caused substantial shifts in GMM estimates but had minimal impact on MLE and regression results. These outliers were identified via a simple leave-one-out approach, and their removal yielded improved agreement across estimators. Developing formalized outlier diagnostics and robust estimation techniques specific to misclassified count data remains an important avenue for future work.

These findings also underscore the importance of modeling error asymmetry. In the ORF example, estimated true positive rates were consistently high and near unity, while true negative rates were substantially lower. Ignoring such asymmetries may have limited practical impact for one parameter but could meaningfully distort inferences for the other. Future studies should explore the real-world implications of these disparities, particularly in contexts such as educational assessment, clinical screening, and behavioral monitoring, where count-based performance measures are common.

Several model extensions warrant further development. Beta-binomial convolution models, already used in our simulation study to represent a mis-specified framework, offer one promising direction for handling overdispersion. More broadly, contamination models with heavy-tailed or outlier-prone error components could improve performance in small-sample or high-variability settings. Finally, robust estimators tailored to discrete misclassification structures are needed to reduce sensitivity to individual data points.

Overall, the proposed framework bridges a critical gap between binary misclassification and continuous error models. While no single estimator dominates in all scenarios, the binomial convolution model and associated inference procedures offer a principled approach to analyzing misclassified count data, with immediate relevance to domains where precision in discrete measurements is paramount.

\FloatBarrier
\onehalfspacing
\bibliographystyle{apalike}
\bibliography{ORF_ME}

\doublespacing
\section{Appendix: Variance Ratio Simulation Results} \label{sec:var_ratio_tables}

\renewcommand{\thetable}{\Roman{table}}
\setcounter{table}{0}

This appendix provides expanded numerical results corresponding to the simulation study reported in Section~\ref{sec:sim_se}, which evaluated the performance of various standard error estimation methods under a correctly specified binomial convolution model. That study used a fully crossed factorial design with 32 distinct simulation settings, each replicated 1{,}000 times. For each setting and estimator, standard errors were computed using both asymptotic (plug-in) and bootstrap-based approaches, and variance ratios were used to assess bias.

Tables~\ref{tab:sim_settings_var_ratios} through \ref{tab:var_ratio_mle_tn} present the full set of results. Table~\ref{tab:sim_settings_var_ratios} summarizes the simulation configurations used. The remaining tables disaggregate variance ratio results by estimator and target parameter: Tables~\ref{tab:var_ratio_reg_tp} and \ref{tab:var_ratio_reg_tn} report results for the regression estimator, one table each for $\pi_{\mathrm{tp}}$ and $\pi_{\mathrm{tn}}$, respectively. Similarly, Tables~\ref{tab:var_ratio_gmm_tp} and \ref{tab:var_ratio_gmm_tn} present results for the GMM estimator, and Tables~\ref{tab:var_ratio_mle_tp} and \ref{tab:var_ratio_mle_tn} provide results for the MLE.

Each variance ratio is defined as
$\mathrm{Ratio}_{\text{method}} = {\mathrm{AV}_{\text{method}}}/{\mathrm{EV}_{\text{mc}}}$,
where $\mathrm{AV}_{\text{method}}$ is the average of the variance estimates produced by the method across the 1{,}000 simulated datasets, and $\mathrm{EV}_{\text{mc}}$ is the empirical Monte Carlo variance of the point estimates over those same datasets.

A ratio near 1 indicates approximate unbiasedness of the variance estimator. Ratios consistently greater than 1 suggest overestimation of variance, while ratios below 1 reflect underestimation. These tables complement the boxplots in Figure~\ref{variance_ratio} by providing detailed numerical summaries of each method’s performance under the full range of simulated conditions.

\begin{table}[!ht]
\small
    \centering
    \caption{Simulation setting for variance ratios}
    \label{tab:sim_settings_var_ratios}
    \begin{tabular}{cccccc}
    \toprule
        $n$ & $N$ & $p$ & $\rho$ & $\pi_{\mathrm{tp}}$ & $\pi_{\mathrm{tn}}$ \\ \hline
        50 & 44 & 0.96 & 0 & 0.98 & 0.75 \\ 
        50 & 69 & 0.96 & 0 & 0.98 & 0.75 \\ 
        50 & 44 & 0.98 & 0 & 0.98 & 0.75 \\ 
        50 & 69 & 0.98 & 0 & 0.98 & 0.75 \\ 
        50 & 44 & 0.96 & 0.03 & 0.98 & 0.75 \\ 
        50 & 69 & 0.96 & 0.03 & 0.98 & 0.75 \\ 
        50 & 44 & 0.98 & 0.03 & 0.98 & 0.75 \\ 
        50 & 69 & 0.98 & 0.03 & 0.98 & 0.75 \\ 
        50 & 44 & 0.96 & 0 & 0.999 & 0.75 \\ 
        50 & 69 & 0.96 & 0 & 0.999 & 0.75 \\ 
        50 & 44 & 0.98 & 0 & 0.999 & 0.75 \\ 
        50 & 69 & 0.98 & 0 & 0.999 & 0.75 \\ 
        50 & 44 & 0.96 & 0.03 & 0.999 & 0.75 \\ 
        50 & 69 & 0.96 & 0.03 & 0.999 & 0.75 \\ 
        50 & 44 & 0.98 & 0.03 & 0.999 & 0.75 \\ 
        50 & 69 & 0.98 & 0.03 & 0.999 & 0.75 \\ 
        50 & 44 & 0.96 & 0 & 0.98 & 0.85 \\ 
        50 & 69 & 0.96 & 0 & 0.98 & 0.85 \\ 
        50 & 44 & 0.98 & 0 & 0.98 & 0.85 \\ 
        50 & 69 & 0.98 & 0 & 0.98 & 0.85 \\ 
        50 & 44 & 0.96 & 0.03 & 0.98 & 0.85 \\ 
        50 & 69 & 0.96 & 0.03 & 0.98 & 0.85 \\ 
        50 & 44 & 0.98 & 0.03 & 0.98 & 0.85 \\ 
        50 & 69 & 0.98 & 0.03 & 0.98 & 0.85 \\ 
        50 & 44 & 0.96 & 0 & 0.999 & 0.85 \\ 
        50 & 69 & 0.96 & 0 & 0.999 & 0.85 \\ 
        50 & 44 & 0.98 & 0 & 0.999 & 0.85 \\ 
        50 & 69 & 0.98 & 0 & 0.999 & 0.85 \\ 
        50 & 44 & 0.96 & 0.03 & 0.999 & 0.85 \\ 
        50 & 69 & 0.96 & 0.03 & 0.999 & 0.85 \\ 
        50 & 44 & 0.98 & 0.03 & 0.999 & 0.85 \\ 
        50 & 69 & 0.98 & 0.03 & 0.999 & 0.85 \\ 
        \bottomrule
    \end{tabular}
\end{table}

\begin{table}[!ht]
\small
    \centering
    \caption{Variance Ratios for Regression Method for True Positive Rate $\pi_{\mathrm{tp}}$} 
    \label{tab:var_ratio_reg_tp}
    \begin{tabular}{ccccc}
    \toprule
        Plug-in & Semipar Boot & Boot & $2\sqrt{n}$ Boot & 2n/3 Boot \\ \hline
        0.9659 & 0.9990 & 0.9797 & 0.9825 & 0.9783 \\ 
        1.0765 & 1.1261 & 1.0694 & 1.0984 & 1.0867 \\ 
        1.0374 & 1.0672 & 1.0468 & 1.0626 & 1.0550 \\ 
        1.0321 & 1.0791 & 1.0203 & 1.0595 & 1.0162 \\ 
        1.0510 & 1.0768 & 1.0420 & 1.0434 & 1.0329 \\ 
        0.9604 & 0.9919 & 0.9456 & 0.9728 & 0.9451 \\ 
        0.9463 & 0.9632 & 0.9294 & 0.9264 & 0.9223 \\ 
        0.9443 & 0.9707 & 0.9237 & 0.9279 & 0.9201 \\ 
        2.1958 & 1.4215 & 1.0145 & 1.0073 & 0.9761 \\ 
        2.0045 & 1.2910 & 0.9242 & 0.9112 & 0.8998 \\ 
        1.5716 & 1.1906 & 0.9017 & 0.8482 & 0.8786 \\
        1.4526 & 1.1091 & 0.8531 & 0.7933 & 0.8189 \\ 
        1.8343 & 1.1951 & 0.8587 & 0.7481 & 0.8139 \\ 
        1.7060 & 1.1158 & 0.8582 & 0.7529 & 0.8166 \\ 
        1.5128 & 1.1078 & 0.8636 & 0.7350 & 0.8202 \\ 
        1.4234 & 1.0282 & 0.8671 & 0.7050 & 0.8042 \\ 
        0.9744 & 1.0148 & 0.9698 & 0.9702 & 0.9718 \\ 
        0.8932 & 0.9345 & 0.9009 & 0.9136 & 0.9017 \\ 
        1.0091 & 1.0302 & 0.9957 & 1.0024 & 0.9853 \\ 
        1.0291 & 1.0635 & 1.0438 & 1.0697 & 1.0432 \\ 
        1.0535 & 1.0847 & 1.0373 & 1.0415 & 1.0453 \\ 
        0.9186 & 0.9513 & 0.9123 & 0.9293 & 0.9106 \\ 
        0.9539 & 0.9727 & 0.9437 & 0.9246 & 0.9291 \\
        1.0290 & 1.0465 & 1.0171 & 1.0276 & 1.0129 \\ 
        1.8896 & 1.3221 & 0.9097 & 0.8889 & 0.8708 \\ 
        1.9251 & 1.3514 & 0.9322 & 0.8962 & 0.9034 \\ 
        1.4870 & 1.1844 & 0.9582 & 0.8752 & 0.9059 \\ 
        1.4864 & 1.1854 & 0.9482 & 0.8584 & 0.9141 \\ 
        1.7287 & 1.1467 & 0.8807 & 0.7604 & 0.8247 \\ 
        1.6791 & 1.1641 & 0.8772 & 0.7674 & 0.8386 \\ 
        1.3722 & 1.0583 & 0.8536 & 0.7387 & 0.8266 \\ 
        1.3373 & 1.0604 & 0.8504 & 0.7200 & 0.8105 \\ \bottomrule
    \end{tabular}
\end{table}

\begin{table}[!ht]
\small
    \centering
    \caption{Variance Ratios for Regression Method for True Negative Rate $\pi_{\mathrm{tn}}$}
    \label{tab:var_ratio_reg_tn}
    \begin{tabular}{ccccc}
    \toprule
        Plug-in & Semipar Boot & Boot & $2\sqrt{n}$ Boot & 2n/3 Boot \\ \hline
        0.9829 & 0.9537 & 0.8719 & 0.8121 & 0.8603 \\ 
        1.0653 & 1.0519 & 0.9422 & 0.8816 & 0.9274 \\ 
        1.0422 & 0.9083 & 0.8242 & 0.7518 & 0.8056 \\ 
        1.0767 & 0.9693 & 0.8573 & 0.7862 & 0.8235 \\ 
        0.9332 & 0.9608 & 0.8757 & 0.8888 & 0.8747 \\ 
        0.9329 & 0.9948 & 0.8796 & 0.9421 & 0.8944 \\ 
        1.0636 & 1.0533 & 0.9265 & 0.9911 & 0.9239 \\ 
        1.0266 & 1.0664 & 0.9355 & 1.0312 & 0.9502 \\ 
        1.0557 & 1.0779 & 0.9304 & 0.9381 & 0.9245 \\ 
        0.9338 & 0.9647 & 0.8643 & 0.8649 & 0.8584 \\ 
        0.9848 & 0.9920 & 0.8525 & 0.8449 & 0.8582 \\ 
        0.9024 & 0.9272 & 0.8172 & 0.7938 & 0.8068 \\ 
        0.9262 & 0.9385 & 0.8040 & 0.8352 & 0.8119 \\ 
        1.0325 & 1.0334 & 0.9307 & 0.9566 & 0.9201 \\ 
        0.9997 & 1.0152 & 0.8620 & 0.9327 & 0.8814 \\ 
        0.8833 & 0.8979 & 0.7873 & 0.8352 & 0.7920 \\ 
        1.1182 & 0.9719 & 0.8496 & 0.7716 & 0.8282 \\ 
        1.0977 & 0.9449 & 0.8323 & 0.7654 & 0.8072 \\ 
        1.3819 & 1.0664 & 0.8826 & 0.8480 & 0.8574 \\ 
        1.3283 & 1.0207 & 0.8822 & 0.8255 & 0.8496 \\ 
        1.1107 & 1.0580 & 0.9061 & 0.8933 & 0.8931 \\ 
        1.0131 & 0.9965 & 0.8871 & 0.8688 & 0.8752 \\ 
        1.1788 & 1.0329 & 0.8374 & 0.9545 & 0.8474 \\ 
        1.1197 & 1.0313 & 0.8765 & 0.9692 & 0.8833 \\ 
        1.0830 & 1.1005 & 0.9302 & 0.8943 & 0.9097 \\ 
        0.9654 & 0.9927 & 0.8499 & 0.8048 & 0.8257 \\ 
        0.9944 & 0.9945 & 0.8453 & 0.8034 & 0.8170 \\ 
        1.0141 & 1.0219 & 0.8998 & 0.8308 & 0.8861 \\ 
        0.9463 & 0.9413 & 0.8126 & 0.8113 & 0.8074 \\ 
        0.9893 & 1.0130 & 0.8884 & 0.8913 & 0.8986 \\ 
        0.8503 & 0.8603 & 0.7236 & 0.7779 & 0.7459 \\ 
        0.9764 & 0.9886 & 0.8161 & 0.8398 & 0.8328 \\ \bottomrule
    \end{tabular}
\end{table}

\begin{table}[!ht]
\small
    \centering
    \caption{Variance Ratios for GMM for True Positive Rate $\pi_{\mathrm{tp}}$}
    \label{tab:var_ratio_gmm_tp}
    \begin{tabular}{ccccc}
    \toprule
        Plug-in & Semipar Boot & Boot & $2\sqrt{n}$ Boot & 2n/3 Boot \\ \hline
        0.3835 & 1.5266 & 1.5072 & 1.4959 & 1.4972 \\ 
        0.4356 & 1.9680 & 1.9232 & 1.9358 & 1.9245 \\ 
        0.4448 & 1.3725 & 1.4050 & 1.3981 & 1.3917 \\ 
        0.4145 & 1.3333 & 1.3077 & 1.3495 & 1.3098 \\ 
        0.4168 & 1.2411 & 1.2437 & 1.2145 & 1.2216 \\ 
        0.4074 & 1.1715 & 1.1439 & 1.1640 & 1.1549 \\ 
        0.3939 & 0.9749 & 0.9971 & 0.9806 & 0.9901 \\ 
        0.3749 & 0.9410 & 0.9256 & 0.9222 & 0.9168 \\ 
        0.5938 & 3.2248 & 3.1414 & 3.0559 & 3.0746 \\ 
        0.5612 & 3.4114 & 3.2777 & 3.0740 & 3.0944 \\ 
        0.9863 & 1.4189 & 1.4023 & 1.2677 & 1.3333 \\ 
        0.4414 & 2.0758 & 2.1642 & 1.9860 & 2.0933 \\ 
        0.5480 & 1.8759 & 1.7674 & 1.5517 & 1.6749 \\ 
        0.5472 & 1.6680 & 1.6233 & 1.3989 & 1.5261 \\ 
        0.5088 & 1.2637 & 1.3226 & 1.0984 & 1.2325 \\ 
        0.4636 & 1.2707 & 1.3887 & 1.1173 & 1.3116 \\ 
        0.4202 & 1.9308 & 1.8871 & 1.9052 & 1.8726 \\ 
        0.3980 & 2.0482 & 2.0181 & 2.0499 & 2.0060 \\ 
        0.3961 & 1.2944 & 1.2988 & 1.3027 & 1.2681 \\ 
        0.4540 & 1.6030 & 1.5923 & 1.6247 & 1.5688 \\ 
        0.4219 & 1.2736 & 1.2796 & 1.2864 & 1.2801 \\ 
        0.3850 & 1.1449 & 1.1348 & 1.1593 & 1.1230 \\ 
        0.3850 & 1.0037 & 1.0176 & 1.0053 & 1.0000 \\ 
        0.4287 & 1.0858 & 1.0770 & 1.0800 & 1.0795 \\ 
        1.1318 & 1.8074 & 1.7738 & 1.7094 & 1.7154 \\ 
        0.3444 & 2.8187 & 2.8268 & 2.6212 & 2.7222 \\ 
        1.3508 & 0.7911 & 0.8051 & 0.7550 & 0.7715 \\ 
        0.4459 & 2.1062 & 2.2105 & 2.0092 & 2.1300 \\ 
        0.5605 & 1.8232 & 1.8373 & 1.5713 & 1.7167 \\ 
        0.4855 & 1.8680 & 1.8221 & 1.6119 & 1.7737 \\ 
        0.9751 & 1.0482 & 1.0577 & 0.9390 & 1.0228 \\ 
        0.4341 & 1.3359 & 1.4886 & 1.2494 & 1.4010 \\ \bottomrule
    \end{tabular}
\end{table}

\begin{table}[!ht]
\small
    \centering
    \caption{Variance Ratios for GMM for True Negative Rate $\pi_{\mathrm{tn}}$}
    \label{tab:var_ratio_gmm_tn}
    \begin{tabular}{ccccc}
    \toprule
        Plug-in & Semipar Boot & Boot & $2\sqrt{n}$ Boot & 2n/3 Boot \\ \hline
        0.3992 & 1.6940 & 1.5363 & 1.3741 & 1.4865 \\ 
        0.4232 & 1.9865 & 1.8096 & 1.6258 & 1.7651 \\ 
        0.4200 & 1.3014 & 1.1915 & 0.9945 & 1.1412 \\ 
        0.3974 & 1.2850 & 1.1653 & 0.9782 & 1.1147 \\ 
        0.3843 & 1.1380 & 1.0790 & 1.0646 & 1.0638 \\ 
        0.3873 & 1.1943 & 1.0918 & 1.1148 & 1.0963 \\ 
        0.4397 & 1.0631 & 0.9671 & 0.9364 & 0.9608 \\ 
        0.4041 & 0.9884 & 0.9175 & 0.9338 & 0.9309 \\ 
        0.4473 & 1.9661 & 1.8110 & 1.8178 & 1.8239 \\ 
        0.4233 & 2.4726 & 2.3417 & 2.2792 & 2.2709 \\ 
        0.5225 & 1.2557 & 1.1320 & 1.0936 & 1.1126 \\ 
        0.4221 & 1.6493 & 1.5565 & 1.5090 & 1.5643 \\ 
        0.3913 & 1.2490 & 1.1204 & 1.1437 & 1.1304 \\ 
        0.3968 & 1.3455 & 1.2278 & 1.2652 & 1.2406 \\ 
        0.3528 & 0.8905 & 0.7986 & 0.8508 & 0.8073 \\ 
        0.3251 & 0.9795 & 0.8941 & 0.9448 & 0.9030 \\ 
        0.4359 & 2.0277 & 1.8581 & 1.6612 & 1.7672 \\ 
        0.4319 & 2.1738 & 1.9775 & 1.7696 & 1.8858 \\ 
        0.4860 & 1.4494 & 1.3290 & 1.1563 & 1.2577 \\ 
        0.5018 & 1.5802 & 1.4560 & 1.2598 & 1.3548 \\ 
        0.4139 & 1.3795 & 1.1986 & 1.1921 & 1.1935 \\ 
        0.3940 & 1.2459 & 1.1546 & 1.1216 & 1.1332 \\ 
        0.4299 & 1.0299 & 0.9301 & 0.9578 & 0.9264 \\ 
        0.4564 & 1.0837 & 0.9692 & 1.0279 & 0.9816 \\ 
        0.5997 & 1.9572 & 1.7746 & 1.7076 & 1.7589 \\ 
        0.4018 & 2.6995 & 2.5516 & 2.3492 & 2.4795 \\ 
        0.6637 & 1.0093 & 0.9354 & 0.9011 & 0.9137 \\ 
        0.4089 & 1.7651 & 1.7233 & 1.6020 & 1.6957 \\ 
        0.4157 & 1.4524 & 1.2930 & 1.3062 & 1.2820 \\ 
        0.4071 & 1.4948 & 1.3622 & 1.3912 & 1.3752 \\ 
        0.4506 & 0.9976 & 0.8806 & 0.9399 & 0.8899 \\ 
        0.3411 & 1.1879 & 1.0927 & 1.1051 & 1.0821 \\ \bottomrule
    \end{tabular}
\end{table}

\begin{table}[!ht]
\small
    \centering
    \caption{Variance Ratios for MLE for True Positive Rate $\pi_{\mathrm{tp}}$}
    \label{tab:var_ratio_mle_tp}
    \begin{tabular}{ccccc}
    \toprule
        Plug-in & Semipar Boot & Boot & $2\sqrt{n}$ Boot & 2n/3 Boot \\ \hline
        0.9785 & 0.9430 & 0.9931 & 0.9642 & 0.9501 \\ 
        1.0919 & 1.0810 & 1.1182 & 1.0827 & 1.0858 \\ 
        1.1055 & 1.0858 & 1.1369 & 1.0867 & 1.0970 \\ 
        1.0671 & 1.0362 & 1.0773 & 1.0428 & 1.0763 \\ 
        1.0779 & 1.0481 & 1.0377 & 0.9991 & 1.0153 \\ 
        1.0202 & 1.0093 & 0.9731 & 0.9453 & 0.9718 \\ 
        0.9660 & 0.9605 & 0.9459 & 0.8889 & 0.9290 \\ 
        0.9463 & 0.9440 & 0.9215 & 0.8812 & 0.8948 \\ 
        1.1856 & 1.0226 & 1.0125 & 1.4217 & 1.0187 \\ 
        1.2589 & 1.1527 & 1.1705 & 1.2777 & 1.0563 \\ 
        1.0632 & 0.9560 & 0.9286 & 0.9122 & 0.9094 \\ 
        1.1390 & 1.0403 & 1.0220 & 1.0809 & 1.0071 \\ 
        1.2155 & 1.0639 & 1.0472 & 0.9959 & 1.0347 \\ 
        1.0937 & 0.9977 & 0.9691 & 0.9363 & 0.9570 \\ 
        1.1325 & 1.0278 & 1.0073 & 0.9654 & 0.9818 \\ 
        1.0658 & 0.9760 & 0.9432 & 0.8886 & 0.9321 \\ 
        1.0609 & 1.0249 & 1.0100 & 1.0279 & 1.0206 \\ 
        0.9898 & 0.9517 & 0.9491 & 0.9676 & 0.9325 \\ 
        1.0654 & 1.0530 & 1.0453 & 1.0295 & 1.0235 \\ 
        1.1720 & 1.1471 & 1.1504 & 1.1774 & 1.1174 \\ 
        1.0841 & 1.0695 & 1.0669 & 1.0089 & 1.0555 \\ 
        0.9464 & 0.9265 & 0.9107 & 0.8627 & 0.8906 \\ 
        1.0068 & 0.9859 & 0.9772 & 0.9277 & 0.9572 \\ 
        1.0416 & 1.0215 & 1.0215 & 0.9968 & 0.9942 \\ 
        1.0208 & 0.9253 & 0.8887 & 0.9627 & 0.8677 \\ 
        1.0484 & 0.9509 & 0.9282 & 0.8888 & 0.9158 \\ 
        1.0925 & 0.9964 & 0.9662 & 0.9283 & 0.9437 \\ 
        1.1063 & 1.0299 & 0.9921 & 0.9728 & 0.9835 \\ 
        1.1147 & 1.0015 & 1.0089 & 0.9436 & 0.9614 \\ 
        1.1215 & 1.0204 & 0.9955 & 0.9431 & 0.9841 \\ 
        1.0866 & 0.9959 & 0.9820 & 0.9201 & 0.9517 \\ 
        1.1534 & 1.0706 & 1.0500 & 0.9818 & 1.0260 \\ \bottomrule
    \end{tabular}
\end{table}

\begin{table}[!ht]
\small
    \centering
    \caption{Variance Ratios for MLE for True Negative Rate $\pi_{\mathrm{tn}}$}
    \label{tab:var_ratio_mle_tn}
    \begin{tabular}{ccccc}
    \toprule
        Plug-in & Semipar Boot & Boot & $2\sqrt{n}$ Boot & 2n/3 Boot \\ \hline
        0.9903 & 0.9911 & 1.0394 & 1.0793 & 1.0298 \\ 
        1.0371 & 1.0554 & 1.1062 & 1.1095 & 1.0807 \\ 
        0.9439 & 0.9966 & 1.1038 & 1.1090 & 1.0660 \\ 
        0.9270 & 0.9437 & 1.0163 & 1.0359 & 1.0256 \\ 
        0.9609 & 0.9926 & 0.9827 & 1.0730 & 1.0008 \\ 
        1.0072 & 1.0582 & 1.0154 & 1.1205 & 1.0440 \\ 
        0.9897 & 1.1113 & 1.0961 & 1.2359 & 1.1274 \\ 
        0.9514 & 1.0767 & 1.0410 & 1.2395 & 1.0926 \\ 
        1.0041 & 1.0328 & 1.0219 & 1.0777 & 1.0042 \\ 
        0.9739 & 0.9880 & 0.9965 & 0.9863 & 0.9596 \\ 
        0.9345 & 0.9947 & 0.9735 & 0.9897 & 0.9755 \\ 
        0.9732 & 1.0119 & 0.9834 & 1.0021 & 0.9822 \\ 
        0.9348 & 0.9815 & 0.9549 & 0.9816 & 0.9506 \\ 
        0.9548 & 0.9849 & 0.9605 & 0.9754 & 0.9494 \\ 
        0.9120 & 1.0065 & 0.9798 & 1.1266 & 0.9993 \\ 
        0.8447 & 0.9081 & 0.8748 & 0.9627 & 0.8978 \\ 
        1.0235 & 0.9947 & 0.9918 & 1.1217 & 1.0256 \\ 
        1.0214 & 0.9885 & 1.0036 & 1.1043 & 0.9999 \\ 
        1.0744 & 1.0946 & 1.1523 & 1.2816 & 1.1485 \\ 
        1.0294 & 1.0490 & 1.0935 & 1.1904 & 1.0741 \\ 
        1.0396 & 1.0499 & 1.0381 & 1.1248 & 1.0424 \\ 
        0.9863 & 1.0172 & 0.9914 & 1.0620 & 1.0002 \\ 
        0.9805 & 1.0572 & 1.0422 & 1.2293 & 1.0883 \\ 
        0.9133 & 1.0095 & 0.9587 & 1.1221 & 1.0025 \\ 
        0.9717 & 1.0059 & 0.9782 & 0.9892 & 0.9628 \\ 
        0.9821 & 0.9901 & 0.9818 & 0.9439 & 0.9766 \\ 
        0.9227 & 0.9682 & 0.9483 & 0.9555 & 0.9570 \\ 
        0.9487 & 0.9844 & 0.9635 & 0.9478 & 0.9507 \\ 
        1.0222 & 1.0595 & 1.0281 & 1.0683 & 1.0231 \\ 
        0.9530 & 0.9762 & 0.9488 & 0.9700 & 0.9687 \\ 
        0.8803 & 0.9726 & 0.9353 & 1.0860 & 0.9715 \\ 
        0.9469 & 1.0144 & 0.9580 & 1.0496 & 0.9511 \\ \bottomrule
    \end{tabular}
\end{table}

\end{document}